\documentclass[12pt,fleqn]{article}
  
\usepackage{latexsym}
\usepackage{amssymb}
\usepackage{amsfonts}
\usepackage{xcolor}
\usepackage{layout}
\usepackage{graphicx}
\usepackage{slashed}
\usepackage{mathtools}
\usepackage{hyperref}
\usepackage{cite}

\newcommand{\nn}{\nonumber}

\begin{document}

\begin{center}
\Large{\bf A toolkit for twisted chiral superfields}\\
\vspace{1cm}

\large Nana Cabo Bizet \footnote{e-mail address: {\tt nana@fisica.ugto.mx}} and Roberto Santos-Silva \footnote{e-mail address: {\tt roberto.santos@academicos.udg.mx}}
\\
[4mm]

{\small \em Departamento de F\'isica, Divisi\'on de Ciencias e Ingenier\'{\i}as, \\ 
Universidad de Guanajuato, Loma del Bosque 103, \\ C.P. 37150, Le\'on, Guanajuato, M\'exico.}\\
[4mm]
{\small \em Departamento Ciencias Naturales y Exactas, CUValles, \\ Universidad de Guadalajara,
Carretera Guadalajara-Ameca Km. 45.5, \\ C.P. 46600, Ameca, Jalisco M\'exico.}\\
[4mm]


\vspace*{2cm}
\small{\bf Abstract}
\end{center}
We calculate the most general terms for arbitrary Lagrangians of twisted chiral superfields in 2D (2,2) supersymmetric theories  \cite{rocek84}. 
The scalar and fermion kinetic terms and interactions are given explicitly. We define a set of twisted superspace coordinates, which allows to obtain Lagrangian terms for generic K\"ahler potential and generic twisted superpotential; this is done in analogy to the corresponding chiral superfields calculations \cite{wessbagger}. 
As examples we obtain the Lagrangian of a single twisted superfield, i.e. the Abelian-dual of the gauged linear sigma model (GLSM) of a single chiral superfield, 
and the Lagrangian for the non-Abelian SU(2) dual of the $\mathbb{CP}^1$ GLSM model. 
Generic Lagrangians contain both twisted-chiral and chiral superfields, with  distinct representations.  
We write down the kinetic terms for all bosons and fermions as well as their interactions for these generic cases. 
As twisted superfields play a central role for T-dualities and Mirror Symmetry in GLSMs, we expect the pedagogical exposition of this 
technique to be useful in those studies.
\begin{center}
\begin{minipage}[h]{14.0cm} {}
\end{minipage}
\end{center}

\bigskip
\bigskip

\date{\today}

\vspace{3cm}

\newpage

\tableofcontents
\section{Introduction}
Twisted chiral representations of 2D Supersymmetry were discovered in the foundational work of Gates, Hull and Rocek \cite{rocek84}. They constitute representations which only appear in two dimensions, because in this case the supersymmetric covariant derivatives
$D_+$ and $\bar D_-$ anticommute. In particular twisted chiral superfields describe field strengths of vector superfields \cite{Gates:1983nr}
and are obtained as T-duals to chiral superfields \cite{Rocek:1991ps}. In theories where chiral and twisted-chiral representations are
present, they are distinguished, otherwise they can be mapped to each other. The infrared limit of a 2D supersymmetric GLSM describe string theory compactified on target spaces, given by the supersymmetric vacua of the GLSM \cite{Witten:1993yc}. Moreover Mirror Symmetry can be obtained
as T-duality   \cite{Strominger:1996it}. For complete intersection Calabi-Yaus described by GLSMs
Mirror Symmetry is realized as T-duality in the GLSM \cite{Strominger:1996it,Morrison:1995yh,Hori:2000kt}. The mirror transformation exchanges
chiral superfields with twisted chiral superfields \cite{Hori:2000kt,MR2003030}. 
These representations of $\mathcal{N}=2$ 2D supersymmetry and their differences with chiral superfield representations
and also Lagrangians with both kinds of multiplets and the arising geometries were studied in \cite{rocek84}. In this work we introduce a calculational technique that serves to explore generic twisted chiral superfields Lagrangians;
as well as Lagrangians with both twisted-chiral and chiral representations. We apply it to determine all
the bosonic and fermionic kinetic terms and interactions of a generic action.


Dimensional reduction links  $\mathcal{N}=1$ supersymmetry in 4D to  (2,2) supersymmetry in 2D \cite{rocek84,Witten:1993yc}
allowing to use the familiar 4D formalism for the later. 
Inspired by the 4D calculations to obtain Lagrangians for the chiral and anti-chiral superfields in general K\"aehler geometry, 
developed by Wess and Bagger \cite{wessbagger}, we perform computations for the Lagrangian  of twisted (anti-)chiral representations. We adopt a new set of superspace coordinates, defining a vector of twisted Grassman coordinates. This set allows a redefinition of the space-time coordinates to 
new variables $\tilde X^{\mu}$  that  are annihilated by covariant derivatives  $D_+$ and $\bar D_-$. With those at hand 
we can write the more generic twisted (anti-)chiral superfields expansions.
Then we can construct polynomials with generic powers of twisted (anti-)chiral fields;
these will serve as building blocks for the twisted superpotential and the K\"ahler potential,
which describe generic Lagrangians. We obtain all bosonic and fermionic contributions to the action including the auxiliary field terms. 
The goal of these notes is to compile a set of clear calculations that 
could serve for further studies. Our key interest are sigma model
Lagrangians arising as non-Abelian T-duals of GLSMs with chiral superfields. In particular there have been great developments in
supersymmetric localization for (2,2) GLSMs, where the partition function on the sphere $S^2$ of a given
GLSM and its Abelian T-dual are matched \cite{gomis1,doroud1,benini1,doroud2,Maxfield:2019czc}.
Our techniques could serve to explore non-Abelian T-duals partition functions, whose
Lagrangians depend on twisted chiral superfields. The work in 2D GLMs in terms of chiral superfields, giving Calabi-Yau target spaces is vast.
In this scheme several dualities have been studied \cite{Hori:2006dk,Hori:2011pd,Gerhardus:2015sla,Knapp:2019cih}. Also there are relevant investigations
in GLSMs with supersymmetric target spaces beyond complete intersection Calabi-Yaus \cite{Caldararu:2007tc,Hosono:2007vf, Jockers:2012zr,Kanazawa:2012xya,Caldararu:2017usq,Hori:2003ic,Gu:2018fpm,Chen:2018wep}. In particular \cite{Caldeira:2018ynv} studies mixed Lagrangians  with both chiral and twisted chiral superfields, which can be explored with the techniques we develop.

The paper is organized as follows: 
In Section \ref{three} we define a set of twisted superspace coordinates and we write the expansions of twisted chiral superfields. In Section
\ref{four} we obtain the general kinetic and interaction terms for Lagrangians depending on twisted chiral superfields,
including boson and fermion contributions.
In Section \ref{five} we calculate an explicit example of Lagrangian, this is the building block of the Mirror Symmetry
description on GLSMs developed by Hori and Vafa \cite{Hori:2000kt}; we also
apply the technique to write the Lagrangian for the $SU(2)$ T-dual of the $\mathbb{CP}^1$ GLSM
model obtained in \cite{CaboBizet:2017fzc}. In Section \ref{six} we calculate arbitrary supersymmetric Lagrangians,
which K\"ahler potential depends on twisted-chiral and chiral superfields. In the last Section \ref{seven}
we present our final comments. In Appendix \ref{one} we give our notation and conventions. In 
Appendix \ref{two} we collect all the terms with fermionic fields from the Lagrangian from Section \ref{six}.

\section{A new set of coordinates and twisted chiral expansions}
\label{three}
In this Section we define a set of twisted superspace coordinates $\tilde \theta$ and a variable change for space-time coordinates  $\tilde X^{m}$. We work
with (2,2) supersymmetry in 2D employing the language of $\mathcal{N}=1$ supersymmetry in 4D.  The 
new variables for the space-time coordinates aid in the derivation of twisted superfield expansions, because they satisfy the relations
$\bar D_+ \tilde{X}^m= D_- \tilde{X}^m=0$. This allows to establish an analogy with the chiral 
superfields expansions. We write the expansions for the twisted chiral and anti-chiral superfields 
in those superspace coordinates. 
The main idea is to exploit the intuition of chiral superfields in global supersymmetry to twisted chiral superfields,
employing the symmetry among those representations.

Let us write next the constraints for twisted chiral and twisted anti-chiral superfields. Twisted superfields satisfy:
\begin{equation}
\label{twc}
\bar{D}_{+} \Psi = D_{-} \Psi= 0,
\end{equation}
and twisted anti-chiral superfields fullfil:
\begin{equation}
\label{atwc}
D_+ \bar\Psi= \bar D _- \bar\Psi=0.
\end{equation}

We search for space-time coordinates redefinitions, involving superspace coordinates, which satisfy the constraints
on twisted (anti-)chiral superfields. This will allow to write expansions satisfying ($\ref{twc}$) and $(\ref{atwc})$.
Thus we define the holomorphic and anti-holomorphic coordinates that satisfy the constraints as follows:

\begin{eqnarray}
\tilde{X}^m& =& x^m + i \theta^+ \sigma_{+ \dot{+}}^m \bar\theta^{\dot +}+i \bar\theta^{\dot -} \sigma_{- \dot{-}}^m \theta^{-}, \label{coord}\\
\bar D_+ \tilde{X}^m&=& D_-\tilde{X}^m=0,\\ 
\bar{\tilde{X}}^m &=& \bar{x}^m - i \theta^+ \sigma_{+ \dot{+}}^m \bar\theta^{\dot +}-i \bar\theta^{\dot -} \sigma_{- \dot{-}}^m \theta^{-}, \\
D_+ \bar{\tilde{X}}^m&=& \bar D_-\bar{\tilde{X}}^m=0.
\end{eqnarray}
Let us rewrite the Grassman superspace coordinates as:
\begin{eqnarray}
\tilde\theta^\alpha&=&\left(\begin{array}{cc} \tilde\theta^{+} & \tilde\theta^{-} \end{array}\right) = \left(\begin{array}{cc} \theta^{+} & \bar\theta^{\dot{-}} \end{array}\right),\\ \bar{\tilde\theta}^\alpha&=&({\tilde\theta}^{\alpha})^{\dag}=\left(\begin{array}{c} \bar{\tilde\theta}^{\dot +} \\ \bar{\tilde\theta}^{\dot{-}} \end{array}\right) = \left(\begin{array}{c} \bar\theta^{\dot +} \\ \theta^{-} \end{array}\right).
\end{eqnarray}
They fulfill the set of relations
\begin{eqnarray}
\tilde\theta^2&=&\tilde\theta^\alpha \tilde\theta_\alpha=-2 \tilde\theta^+ \tilde\theta^-=-2 \theta^+ \bar\theta^{\dot{-}} , \\
{\bar{\tilde{\theta}}}^2&=&\bar{\tilde{\theta}}_{\dot{\alpha}} \bar{\tilde{\theta}}^{\dot{\alpha}}=2 \bar{\tilde{\theta}}^{\dot{+}} \bar{\tilde{\theta}}^{\dot{-}}=2 \bar\theta^{\dot{+}}\theta^{-},\nonumber \\
d\tilde\theta^2&=&d\tilde\theta^\alpha d \tilde\theta_\alpha=-2 d\tilde\theta^+ d\tilde\theta^-=-2d \theta^+ d\bar\theta^{\dot{-}} , \\
d{\bar{\tilde{\theta}}}^2&=&d\bar{\tilde{\theta}}_{\dot{\alpha}} d\bar{\tilde{\theta}}^{\dot{\alpha}}=2 d\bar{\tilde{\theta}}^{\dot{1}} d\bar{\tilde{\theta}}^{\dot{2}}=2 d\bar\theta^{\dot{+}}d\theta^-.\nonumber
\end{eqnarray}
This implies that integrals on superspace can be expressed as
\begin{eqnarray}
\int {d \theta}^4&=&\frac{1}{4} \int d \theta^+ d \theta^- d \bar\theta^- d \bar\theta^+=-\frac{1}{16}\int {d \tilde\theta}^2 {d \bar{\tilde\theta}}^2,\\
\int d \theta^+ d \theta^- d \bar\theta^- d \bar\theta^+&=& - \int d \tilde \theta^+ d \tilde\theta^- d \bar{\tilde\theta}^- d \bar{\tilde\theta}^+ ,\\
\int {d \theta}^4 {\tilde\theta}^2 {\bar{\tilde{\theta^2}}}&=&-1, \, \int {d \theta}^4 {\theta}^2 {\bar{{\theta^2}}}=1.
\end{eqnarray}
The advantage of the redefinition is that one can now rewrite the superspace redefined coordinates of
(\ref{coord}) in the form:
\begin{eqnarray}
\label{st-tq}
\tilde{X}^m = x^m + i \tilde\theta^\alpha \sigma_{\alpha \dot{\alpha}}^m \bar{\tilde\theta}^{\dot\alpha}, \\ \bar{\tilde{X}}^m = x^m - i \tilde\theta^\alpha \sigma_{\alpha \dot{\alpha}}^m \bar{\tilde\theta}^{\dot\alpha}.
\end{eqnarray}
Therefore, the most general combination that satisfies the twisted chiral superfield constraint can be written as
\begin{equation}
\label{twsf}
\Psi(\tilde X) = \psi(\tilde X) + \sqrt{2} \tilde\theta^{\alpha} \tilde\chi_\alpha (\tilde X) + \tilde\theta^\alpha \tilde\theta_\alpha G(\tilde X).
\end{equation}
Now expanding in Taylor series of the Grassman coordinates around $x^\mu$ we get the expression
\begin{eqnarray}
\Psi (x)&=& \psi(x)+i \tilde\theta^\alpha \sigma^m_{\alpha \dot\alpha} \bar{\tilde\theta}^{\dot\alpha}\partial_m \psi(x) -\frac{1}{2} \tilde\theta^\alpha \sigma^m_{\alpha \dot\alpha} \bar{\tilde\theta}^{\dot\alpha} \tilde\theta^\beta \sigma^n_{\beta \dot\beta} \bar{\tilde\theta}^{\dot\beta} \partial_m \partial_n \psi(x) +\nonumber \\ 
&&\sqrt{2} \tilde\theta^{\alpha}[\tilde\chi_{\alpha}(x)+i \tilde\theta^\beta \sigma^m_{\beta \dot\beta} \bar{\tilde\theta}^{\dot\beta} \partial_m \tilde\chi_\alpha(x)] + \tilde\theta^{\alpha} \tilde\theta_{\alpha} G(x).
\end{eqnarray}
Employing the properties of the Grassman coordinates, and the product of the Pauli matrices, the terms
in previous equation can be written in a compact form as:
\footnote{This is obtained from $-\frac{1}{2} \tilde\theta^\alpha \sigma^m_{\alpha \dot\alpha} \bar{\tilde\theta}^{\dot\alpha} \tilde\theta^\beta \sigma^n_{\beta \dot\beta} \bar{\tilde\theta}^{\dot\beta} \partial_m \partial_n A(x) = -\frac{1}{4}\tilde\theta^2 \bar{\tilde\theta}^2 \square A (x)$ and $i \sqrt{2} \tilde\theta^{\alpha} \tilde\theta^\beta \sigma^m_{\beta \dot\beta} \bar{\tilde\theta}^{\dot\beta} \partial_m \tilde\chi_\alpha(x)=- \frac{i}{\sqrt{2}} \epsilon^{\beta \alpha} \tilde\theta^2 \bar{\tilde\theta}^{\dot\beta} \sigma^m_{\beta \dot\beta} \partial_m \tilde\chi_\alpha(x)= \frac{i}{\sqrt{2}} \tilde\theta^2 \bar{\tilde\theta}^{\dot\alpha} \sigma^m_{\alpha \dot\alpha} \partial_m \tilde\chi^\alpha(x)$.}

\begin{eqnarray}
\label{twsf:expand}
\Psi &=& \psi(x)+ \sqrt{2} \tilde\theta^{\alpha} \chi_{\alpha}(x)+\tilde\theta^2 G(x)+\frac{i}{\sqrt{2}} \tilde\theta^2 \bar{\tilde\theta}^{\dot\alpha} \sigma^m_{\alpha \dot\alpha} \partial_m \tilde\chi^\alpha(x) \nonumber \\
&&i \tilde\theta^\alpha \sigma^m_{\alpha \dot\alpha} \bar{\tilde\theta}^{\dot\alpha}\partial_m \psi(x) -\frac{1}{4}\tilde\theta^2 \bar{\tilde\theta}^2 \square \psi(x).
\end{eqnarray}
Similarly the twisted anti-chiral superfield $\bar\Psi$ can be written in terms of the redefined 
superspace coordinates as
\begin{equation}
\label{atwsf}
\bar\Psi = \bar{\psi}(\bar{\tilde X}) + \sqrt{2} \tilde\theta^{\dot\alpha} \bar{\tilde\chi}_{\dot\alpha} (\bar{\tilde X}) + \bar{\tilde\theta}^{\dot\alpha} \bar{\tilde\theta}_{\dot\alpha} \bar{G}(\bar{\tilde X}).
\end{equation}
Expanding the twisted anti-chiral superfield $\bar\Psi$ we have a similar expression to (\ref{twsf:expand}):
\begin{eqnarray}
\label{atwsf:expand}
\bar\Psi &= &\bar \psi(x)+ \sqrt{2} \bar{\tilde\theta}^{\dot\alpha} \bar{\tilde\chi}_{\dot\alpha}(x)+\bar{\tilde\theta}^2 \bar{G}(x)+\frac{i}{\sqrt{2}} \bar{\tilde\theta}^2 \partial_{m} \bar{\tilde\chi}^{\dot\alpha}(x) \sigma^{m}_{\alpha \dot\alpha} \tilde{\theta}^{\alpha} \nonumber \\ 
&-& i \tilde\theta^\alpha \sigma^{m}_{\alpha \dot\alpha} \bar{\tilde\theta}^{\dot\alpha}\partial_{m} \bar{\psi}(x) -\frac{1}{4}\tilde\theta^2 \bar{\tilde\theta}^2 \square \bar{\psi} (x).
\end{eqnarray}
Let us discuss now standard terms in supersymmetric Lagrangians. 
For generic Lagrangians it will be necessary to compute products of
twisted chiral superfields
\begin{equation}
\Psi^\mu \Psi^\nu = \psi^\mu \psi^\nu + \sqrt{2} \tilde\theta^{\alpha} [\tilde\chi^\nu \psi^\mu+ \tilde\chi^\mu \psi^\nu] + \tilde\theta^\alpha \tilde\theta_\alpha [G^\nu \psi^\mu G^\mu \psi^\nu- \tilde\chi^{\nu \beta} \tilde\chi^\mu_\beta],
\end{equation}
which is also a twisted chiral superfield. Those terms could appear in the twisted superpotential.\\

We also compute products of twisted chiral and twisted anti-chiral superfields
\begin{eqnarray}
\bar\Psi^{\bar\mu} \Psi^{\bar\nu} = \bar \psi^{\bar\mu} \psi^\nu + \sqrt{2} \tilde\theta^\beta \tilde\chi_\beta^\nu \bar{\psi}^{\bar\mu}+ \tilde\theta^\beta \tilde\theta_\beta \bar{G}^{\bar\mu} \psi^\nu + \sqrt{2} \bar{\tilde\theta}^{\dot\alpha} \bar{\tilde\chi}_{\dot\alpha}^{\bar\mu} \psi^{\nu} + 2 \bar{\tilde\theta}^{\dot\alpha} \bar{\tilde\chi}_{\dot\alpha}^{\bar \mu} \tilde\theta^\beta \tilde\chi^{\nu}_\beta \nonumber \\ +\sqrt{2} \bar{\tilde\theta}^{\dot\alpha} \bar{\tilde\chi}_{\dot\alpha}^{\bar\mu} \tilde\theta^\beta \tilde\theta_\beta G^{\nu}+\bar{\tilde\theta}^{\dot\alpha} \bar{\tilde\theta}_{\dot\alpha} \bar{G}^{\bar\mu} \psi^{\nu} + \sqrt{2} \bar{\tilde\theta}^{\dot\alpha} \bar{\tilde\theta}_{\dot\alpha} \tilde\theta^\beta \tilde \chi_\beta^{\nu} \bar{G}^{\bar\mu}+ \bar{\tilde\theta}^2 \tilde\theta^2 \bar{G}^{\bar\mu} G^\nu .
\end{eqnarray}
This of course is not a twisted chiral or anti-chiral superfield. Those terms could appear
in the K\"ahler potential. 

Let us now compute cubic powers of twisted superfields:
\begin{eqnarray}
\Psi^\mu \Psi^\nu \Psi^\rho = \psi^\mu \psi^\nu \psi^\rho + \sqrt{2} \tilde\theta^\alpha [\psi^\mu \psi^\nu \tilde\chi_\alpha^\rho+\tilde\chi_{\alpha}^\nu \psi^\mu \psi^\rho+ \tilde\chi_\alpha^\mu \psi^\nu \psi^\rho] + \nonumber \\ \tilde\theta^2 [\psi^\mu \psi^\nu G^\rho+G^\nu \psi^\mu \psi^\rho +G^\mu \psi^\nu \psi^\rho -\tilde\chi^{\mu \alpha} \tilde\chi_\alpha^\nu \psi^\rho -\tilde\chi^{\nu \alpha} \tilde\chi_{\alpha}^\rho \psi^\nu- \tilde\chi^{\mu \alpha} \tilde\chi_\alpha^\rho \psi^\nu ].
\end{eqnarray}
Note that previous results were computed in the $X$ and $\tilde X$ variables. 

\medskip
Now let us write down the more general 2D (2,2) supersymmetric Lagrangian to order three in powers of the fields using only twisted chiral fields. It is given by:

\begin{eqnarray}
\mathcal{L}= \left. \bar\Psi^{\bar\mu} \Psi^\mu \right\vert_{\tilde\theta^2 \bar{\tilde\theta}^2} + \left( \left. \left(\lambda_\mu \Psi^\mu +\frac{1}{2} m_{\mu \nu} \Psi^\mu \Psi^\nu+\frac{1}{3} g_{\mu \nu \rho}\Psi^\mu \Psi^\nu \Psi^\rho \right) \right\vert_{\tilde\theta^2} + c.c. \right).\label{Lcubic}
\end{eqnarray}
The scalar coefficients are $\lambda_\mu$, $m_{\mu \nu}$ and $g_{\mu \nu \rho}$. The last two are symmetric in their indices. Substituting the respective contributions of the previous expressions (\ref{twsf:expand}) and (\ref{atwsf:expand}) in (\ref{Lcubic}), the Lagrangian becomes

\begin{eqnarray}
\mathcal{L}&=& -\bar{\psi}^{\bar\mu} \square \psi^{\mu} + i \partial_{m} \bar{\tilde\chi}^{\bar\mu \dot\alpha} \sigma_{\alpha \dot\alpha}^{m} \tilde\chi^{\mu \alpha}+\bar{G}^{\bar\mu} G^{\mu}+ \lambda_\mu G^\mu+ \\
&+&\frac{1}{2} m_{\mu \nu} ( G^\nu \psi^\mu+ G^\mu \psi^\nu - \tilde\chi^{ \nu \alpha} \tilde\chi_\alpha^\mu) \nonumber \\ 
&+&\frac{1}{3} g_{\mu \nu \rho}(A^\mu \psi^\nu G^\rho+ \psi^\mu \psi^\rho G^\nu+ \psi^\nu \psi^\rho G^\mu) \nonumber \\
&-&\frac{1}{3} g_{\mu \nu \rho}( \tilde\chi^{\mu \alpha} \tilde\chi^\nu_\alpha \psi^\rho- \tilde\chi^{\nu \alpha} \tilde\chi_\alpha^\rho \psi^\mu - \tilde\chi^{\mu \alpha} \tilde\chi_{\alpha}^\rho \psi^\nu) + c.c..\nonumber
\end{eqnarray}
This expression can be simplified into
\begin{eqnarray}
\mathcal{L}&=& -\bar{\psi}^{\bar\mu} \square \psi^\mu + i \partial_{m} \bar{\tilde\chi}^{\bar\mu \dot\alpha} \sigma_{\alpha \dot\alpha}^{m} \tilde\chi^{\mu \alpha}+\bar{G}^{\bar\mu} G^{\mu}+ \lambda_{\mu} G^{\mu}+ m_{\mu \nu} (G^\mu \psi^\nu - \frac{1}{2}\tilde\chi^{ \mu \alpha} \tilde\chi_\alpha^{\nu}) \nonumber \\& +& g_{\mu \nu \rho} (\psi^{\nu} \psi^{\nu} G^{\rho} - \tilde\chi^{\mu \alpha} \tilde\chi^{\nu}_\alpha \psi^{\rho}) + c.c. 
\end{eqnarray}
It is interesting to compare this result to the Lagrangian obtained for chiral superfields in
\cite{wessbagger}, which is similar up to signs and the twisted components notation.
One could integrate out the auxiliar fields $G$'s in order to obtain just the dynamical variables.
We leave this for next section, where the more general K\"ahler potential and twisted superpotential 
are studied.


\section{K\"ahler Geometry}
\label{four}

In this Section we construct the most general 2D (2,2) supersymmetric Lagrangian depending on twisted (anti-)chiral superfields. 
They are decomposed as series expansions in powers of the superfields. We include
a generic K\"ahler potential and a generic twisted superpotential. The building blocks
will be generic functions ($\bar Q$)$Q$ of twisted (anti-)chiral superfields,  as in the approach of \cite{wessbagger}.

Let us consider $n$ twisted chiral superfields $\Psi^{\mu}, \mu=1...n$ with Lagrangian given by

\begin{equation}
\label{lagrangian1}
\mathcal{L}= \int d^2\tilde\theta d^2\bar{\tilde\theta} K(\Psi^\mu, \bar{\Psi}^\mu)+ \left( \int d^2 \tilde\theta W(\Psi^\mu)+ c.c.\right).
\end{equation}
$K$ is a function that depends on $n$ holomorphic and anti-holomorphic variables $\Psi^\mu$ and $\bar\Psi^\mu$ respectively.
$W$ is a holomorphic function (i.e. only is function of $\Psi^\mu$). Let us expand $K$ in terms of the twisted chiral fields

\begin{equation}
\label{kexpan}
K(\Psi^\mu,\bar\Psi^\mu)=\sum_{i,j}\sum_{\mu_1,\cdots, \mu_i, \nu_1,\cdots,\nu_j} k_{\mu_1\cdots \mu_i \nu_1\cdots\nu_j} \Psi^{\mu_1} \cdots \Psi^{\mu_i}\bar\Psi^{\nu_1} \cdots \bar\Psi^{\nu_j},
\end{equation}
where $k_{\mu_1\cdots \mu_i \nu_1\cdots\nu_j}$ are the constants related to the Taylor expansion. 

In a similar way we expand a generic function of twisted chiral superfields $Q$ as

\begin{equation}
Q(\Psi^\mu)=\sum_i\sum_{\mu_1,\cdots, \mu_i} p_{\mu_1\cdots \mu_i} \Psi^{\mu_1} \cdots \Psi^{\mu_i},
\end{equation}
here $p_{\mu_1\cdots \mu_i}$ are the coefficients associated to the expansion.  $Q$ could represent
the twisted superpotential $W$. First we expand $Q$ and its complex conjugate explicitly around the scalar component $\psi$, in order to use those expressions to compute the K\"ahler potential $K$, we have

\begin{eqnarray}
\label{poly1}
Q(\Psi)&=& Q(\psi) + \sqrt{2} \tilde\theta^{\alpha} \tilde\chi^{\mu}_\alpha \frac{\partial}{\partial \psi^{\mu} } Q + \tilde\theta^2 G^\mu \frac{\partial}{\partial \psi^{\mu} } Q+ \left( \frac{\partial^2}{\partial \psi^\mu \partial \psi^\nu} Q \right) \tilde\theta^{\alpha} \tilde\chi_{\alpha}^\mu \tilde\theta^\beta \tilde\chi^\nu_\beta \nonumber \\ 
&=& Q(\psi) + \sqrt{2} \tilde\theta^{\alpha} \tilde\chi^{\mu}_\alpha \frac{\partial Q}{\partial \psi^{\mu} } + \tilde\theta^2 \left[ G^\mu \frac{\partial Q}{\partial \psi^{\mu} } -\frac{1}{2} \frac{\partial^2 Q}{\partial \psi^\mu \partial \psi^\nu} \tilde\chi^{\alpha \mu} \tilde\chi^\nu_\alpha \right].
\end{eqnarray}
In a similar manner we compute $\bar{Q} (\bar \Psi)$

\begin{equation}
\label{poly2}
\bar Q (\bar\Psi)= \bar Q(\bar \psi) + \sqrt{2} \bar{\tilde\theta}^{\dot \alpha} \bar{\tilde\chi}^{\bar\mu}_{\dot \alpha} \frac{\partial \bar Q}{\partial \bar \psi^{\bar\mu} } + \bar{\tilde\theta}^2 \left[ \bar G^{\bar\mu} \frac{\partial \bar Q}{\partial \bar \psi^{\bar\mu} } -\frac{1}{2} \frac{\partial^2 \bar Q}{\partial \bar \psi^{\bar\mu} \partial \bar \psi^{\bar\nu}} \bar{\tilde\chi}^{\bar\mu}_{\dot\alpha} \bar{\tilde\chi}^{\dot\alpha \bar\nu} \right].
\end{equation}
Now let us compute the component $\tilde\theta^2 \bar{\tilde\theta}^2$ of $K(\Psi^{\mu},\bar\Psi^{\mu})=\Psi^{\mu_1} \cdots \Psi^{\mu_i}\bar\Psi^{\nu_1} \cdots \bar\Psi^{\nu_j}$, using the previous expressions, if we set $Q=\Psi^{\mu_1} \cdots \Psi^{\mu_i}$ and $\bar Q=\bar\Psi^{\nu_1} \cdots \bar\Psi^{\nu_j}$, this implies $Q(A)=\psi^{\mu_1} \cdots \psi^{\mu_i}$ and $\bar Q(\bar \psi)= \bar{\psi}^{\nu_1} \cdots \bar{\psi}^{\nu_j}$. We must bare in mind that these expressions are superspace valued i.e. they depend on $\tilde X^m= x^m + i \tilde \theta^\alpha \sigma_{\alpha \dot\alpha}^m \bar{\tilde \theta}^{\dot \alpha}$ and their c.c., thus 
\begin{equation}
Q(\psi)=\psi^{\mu_1} \cdots \psi^{\mu_i} =\left. \left[ \psi^{\mu_1} \cdots \psi^{\mu_i} \right] \right\vert_{x}+i \left. \left[ \frac{\partial}{\partial x^k} \psi^{\mu_1} \cdots \psi^{\mu_i} \right]\right\vert_{x} \tilde\theta^\alpha \sigma_{\alpha \dot\alpha}^k \bar{\tilde \theta}^{\dot \alpha}.
\end{equation}
A similar expression holds for $\bar Q(\bar \psi)$. Recall that $x^k$ are the space-time coordinates. In a similar fashion we expand the term $\tilde\chi_\alpha^\rho \frac{\partial Q(\psi)}{\partial \psi^\rho}$ in terms of the superspace coordinates, so we arrive to the following equation

\begin{eqnarray}
\tilde\chi_\alpha^\rho \frac{\partial Q(\psi)}{\partial \psi^\rho}= \left. \left[ \tilde\chi_\alpha^\rho \frac{\partial Q(\psi)}{\partial \psi^\rho} \right] \right\vert_x + i \left. \left[ \frac{\partial}{\partial x^k} \tilde\chi_\alpha^\rho Q(\psi)\right] \right\vert_x \tilde\theta^\alpha \sigma_{\alpha \dot\alpha}^k \bar{\tilde \theta}^{\dot \alpha}, \nonumber \\ = \left. \left[ \tilde\chi_\alpha^\rho \frac{\partial}{\partial \psi^\rho} \psi^{\mu_1} \cdots \psi^{\mu_i} \right] \right\vert_x + i \left. \left[ \frac{\partial}{\partial x^k} \psi^{\mu_1} \cdots \psi^{\mu_i} \right] \right\vert_x \tilde\theta^\alpha \sigma_{\alpha \dot\alpha}^k \bar{\tilde \theta}^{\dot \alpha}, 
\end{eqnarray}
and a similar expression holds for $\bar{\tilde\chi}_\alpha^\rho \frac{\partial \bar Q(\bar \psi)}{\partial \bar \psi^\rho}$. 

The expansion of $K$ up to order $\tilde\theta^2 \bar{\tilde\theta}^2$ is given by

\begin{eqnarray}
\left. K_{i j} \right\vert_{\tilde\theta^2 \bar{\tilde\theta}^2} &=& \Big[ \left( G^\rho \frac{\partial}{\partial \psi^\rho} (\psi^{\mu_1} \cdots \psi^{\mu_i}) -\frac{1}{2} {\tilde\chi}^{\rho_1} {\tilde\chi}^{\rho_2} \frac{\partial^2}{\partial \psi^{\rho_1} \partial \psi^{\rho_2} }(\psi^{\mu_1} \cdots \psi^{\mu_i})\right) \cdot \nonumber \\ & \cdot& \left( \bar{G}^\rho \frac{\partial}{\partial \bar{\psi}^\rho} (\bar{\psi}^{\nu_1} \cdots \bar{\psi}^{\nu_j}) -\frac{1}{2} \bar{\tilde\chi}^{\rho_1} \bar{\tilde\chi}^{\rho_2} \frac{\partial^2}{\partial \bar{\psi}^{\rho_1} \partial \bar{\psi}^{\rho_2} }(\bar{\psi}^{\nu_1} \cdots \bar{\psi}^{\nu_j}) \right) \nonumber \\ 
&-&\frac{1}{2} \frac{\partial}{\partial x^i}(\psi^{\mu_1} \cdots \psi^{\mu_i})\frac{\partial}{\partial \bar{x}_i} (\bar{\psi}^{\nu_1} \cdots \bar{\psi}^{\nu_j}) \nonumber \\ 
&-&i \frac{\partial}{\partial \bar{\psi}^{\rho_1}}(\bar{\psi}^{\nu_1} \cdots \bar{\psi}^{\nu_j}) \bar{\tilde\chi}^{\rho_1} \bar\sigma^{j} \frac{\partial}{\partial x^{j}} \left( \tilde\chi^{\rho_2} \frac{\partial}{\partial \psi^{\rho_2}} (\psi^{\mu_1} \cdots \psi^{\mu_i}) \right)\Big].
\end{eqnarray}
This expression is valued on space-time, i.e. it is given in terms of the coordinates $x^m$. Next, we proceed to rearrange this expression in terms of the scalar components $k_{ij}=\psi^{\mu_1} \cdots \psi^{\mu_i} \bar{\psi}^{\nu_1} \cdots \bar{\psi}^{\nu_j}$. The term in the third line can be rearranged
as $\sum_{l,m}-\partial_i \psi^{\mu_l}\partial_i \bar \psi^{\nu_m}\partial_{\mu_l\nu_m} k_{ij}$. Employing similar
substitutions we get

\begin{eqnarray} \label{kahler}
\left. K_{i j} \right\vert_{\tilde\theta^2 \bar{\tilde\theta}^2} &=& \Big[ G^\mu \bar G^\nu \frac{\partial^2}{\partial A^\mu \partial \bar \psi^\nu}k_{ij} -\frac{1}{2} G^\mu \bar{\tilde\chi}^\nu \bar{\tilde\chi}^\rho \frac{\partial^3}{\partial \psi^\mu \partial \bar \psi^\nu \partial \bar \psi^\rho} k_{ij} \nonumber \\ 
&-&\frac{1}{2} \bar{G}^\mu {\tilde\chi}^\nu {\tilde\chi}^\rho \frac{\partial^3}{\partial \bar \psi^\mu \partial \psi^\nu \partial \psi^\rho} k_{ij} +\frac{1}{4} \tilde\chi^\mu {\tilde\chi}^\nu \bar{\tilde\chi}^\rho \bar{\tilde\chi}^\sigma \frac{\partial^4}{\partial \psi^\mu \partial \bar \psi^\nu \partial \bar \psi^\rho \partial \bar \psi^\sigma} k_{ij} \nonumber \\ 
&-&\partial_k \psi^\mu \partial^k \bar \psi^\nu \frac{\partial^2}{\partial \psi^\mu \partial \bar \psi^\nu} k_{ij} -i \bar{\tilde\chi}^\mu \bar\sigma^k \partial_k \chi^\nu \frac{\partial^2}{\partial \psi^\mu \partial \bar \psi^\nu} k_{ij} \nonumber\\ 
&-&i \bar{\tilde\chi}^\sigma \bar\sigma^k \tilde\chi^\mu \partial_k \psi^\nu \frac{\partial^3}{\partial \psi^\mu \partial \psi^\nu \partial \bar \psi^\sigma} k_{ij} \Big].
\end{eqnarray}
In previous formula space-time derivatives are denoted by $\partial_k=\frac{\partial}{\partial x^k}$. Next we sum the terms proportional to $\tilde \theta^2 \bar{\tilde \theta}^2$ over all the expansion $(\ref{kexpan})$. Let us define the following expressions 

\begin{eqnarray}
\label{metric}
g_{\mu \bar\nu}&=&\frac{\partial^2}{\partial \psi^\mu \partial \bar \psi^\nu} k,\,  \,  \,  k=\sum_{ij} k_{ij}, \\ \label{chris}
\frac{\partial}{\partial \psi^\rho} g_{\mu \bar\nu} &=& g_{\sigma \bar\nu} \Gamma^\sigma_{\mu \rho}, \\ \label{chrisc} \frac{\partial}{\partial \bar \psi^\rho} g_{\mu \bar \nu} &=& g_{\mu \bar \sigma} \Gamma^{\bar \sigma}_{\bar\nu \bar\rho}.
\end{eqnarray} 
Using the previous expressions we rewrite $(\ref{kahler})$ as:

\begin{eqnarray} \label{kahler2}
\sum_{ij}\left. K_{i j} \right\vert_{\tilde\theta^2 \bar{\tilde\theta}^2} &=& \Big[ G^\mu \bar G^{\bar\nu} g_{\mu \bar\nu} -\frac{1}{2} G^\mu \bar{\tilde\chi}^{\bar \nu} \bar{\tilde\chi}^{\bar\rho} g_{\mu \bar\sigma} \Gamma^{\bar \sigma}_{\bar \nu \bar \rho} -\frac{1}{2} \bar{G}^{\bar\mu} {\tilde\chi}^\nu {\tilde\chi}^\rho g_{\sigma \bar\mu} \Gamma^{\sigma}_{\nu \rho} \nonumber \\ &+&\frac{1}{4} \tilde\chi^\mu {\tilde\chi}^\nu \bar{\tilde\chi}^{\bar \rho} \bar{\tilde\chi}^{\bar \sigma} \frac{\partial^2}{\partial \psi^\nu \partial \bar \psi^{\bar\sigma} } g_{\mu \bar\rho} -(\partial_k \psi^\mu \partial^k \bar \psi^{\bar\nu}) g_{\mu \bar\nu} -i \bar{\tilde\chi}^{\bar\mu} \bar\sigma^k \partial_k \chi^\nu g_{\nu \bar\mu} \nonumber\\ &-&i \bar{\tilde\chi}^{\bar \sigma} \bar\sigma^k \tilde\chi^\mu \partial_k \psi^\nu g_{\theta \bar \sigma} \Gamma^\theta_{\mu \nu} \Big]. \label{kij1}
\end{eqnarray}
Substituting (\ref{kij1}) into $(\ref{lagrangian1})$ we write down explicitly the Lagrangian. The kinetic term $\int d^2\tilde\theta d^2\bar{\tilde\theta} K(\Psi^\mu, \bar{\Psi}^\mu)$ is given by previous computation. Using expressions $(\ref{poly1})$ and $(\ref{poly2})$ it is easy to compute the twisted superpotential term given explicitly by

\begin{eqnarray}
&&\int \left( d^2 \tilde\theta W(\Psi^\mu)+ d^2 \bar{\tilde\theta} \bar{W} (\bar\Psi^\mu)\right)= \nonumber \\
&&G^\mu \partial_\mu W -\frac{1}{2} (\partial_\mu \partial_\nu W) \tilde\chi^{\alpha \mu} \tilde\chi^\nu_\alpha +\bar G^{\bar\mu} \partial_{\bar\mu} \bar W -\frac{1}{2} (\partial_{\bar\mu} \partial_{\bar\nu} \bar W) \bar{\tilde\chi}^{\bar\mu}_{\dot\alpha} \bar{\tilde\chi}^{\dot\alpha \bar\nu}.
\end{eqnarray}
Finally the Lagrangian can be written as

\begin{eqnarray}
\mathcal{L}&=&G^\mu \bar G^{\bar\nu} g_{\mu \bar\nu} -\frac{1}{2} G^\mu \bar{\tilde\chi}^{\bar\nu} \bar{\tilde\chi}^{\bar\rho} g_{\mu \bar \sigma} \Gamma^{\bar \sigma}_{\bar \nu \bar \rho} -\frac{1}{2} \bar{G}^{\bar\mu} {\tilde\chi}^\nu {\tilde\chi}^\rho g_{\sigma \bar\mu} \Gamma^{\sigma}_{\nu \rho} \nonumber \\ 
&+&\frac{1}{4} \tilde\chi^\mu {\tilde\chi}^{\nu} \bar{\tilde\chi}^{\bar\rho} \bar{\tilde\chi}^{\bar\sigma} \frac{\partial^2}{\partial \psi^\nu \partial \bar \psi^{\bar\sigma} } g_{\mu \bar\rho} -(\partial_m \psi^\mu \partial^m \bar \psi^{\bar\nu}) g_{\mu \bar\nu} -i \bar{\tilde\chi}^{\bar\mu} \bar\sigma^m \partial_m \chi^\nu g_{\nu \bar\mu} \nonumber\\ 
&-&i \bar{\tilde\chi}^{\bar\sigma} \bar\sigma^\rho \tilde\chi^\mu \partial_\rho \psi^\nu g_{\theta \bar \sigma} \Gamma^\theta_{\mu \nu} + G^\mu \partial_\mu W -\frac{1}{2} (\partial_\mu \partial_\nu W) \tilde\chi^{\alpha \mu} \tilde\chi^\nu_\alpha + \nonumber \\ 
&&\bar G^{\bar\mu} \partial_{\bar\mu} \bar W -\frac{1}{2} (\partial_{\bar\mu} \partial_{\bar\nu} \bar W) \bar{\tilde\chi}^{\bar\mu}_{\dot\alpha} \bar{\tilde\chi}^{\dot\alpha \bar\nu}.
\end{eqnarray}
The sixth and seventh terms in previous expression combine as $-i \bar{\tilde\chi}^{\bar\mu} \bar\sigma^m \partial_m \chi^\nu g_{\nu \bar\mu} -i \bar{\tilde\chi}^{\bar\sigma} \bar\sigma^m \tilde\chi^\mu \partial_m \psi^\nu g_{\theta \bar \sigma} \Gamma^\theta_{\mu \nu}=-i \bar{\tilde\chi}^{\bar\mu} \sigma^m g_{\nu \bar\mu} \left( \partial_m \tilde\chi^\nu + \tilde\chi^\sigma \partial_m \psi^\theta \Gamma^\nu_{\sigma \theta} \right)$. Here the term in parenthesis is the covariant derivative

\begin{eqnarray}
D_m\tilde\chi^\nu = \partial_m \tilde\chi^\nu + \tilde\chi^\sigma \partial_m \psi^\theta \Gamma^\nu_{\sigma \theta}.
\end{eqnarray}
Plugin in this expression and rewriting the Lagrangian we get the following expression

\begin{eqnarray}
\mathcal{L}&=&G^\mu \bar G^{\bar\nu} g_{\mu \bar\nu} -\frac{1}{2} G^\mu \left( \bar{\tilde\chi}^{\bar\nu} \bar{\tilde\chi}^{\bar\rho} g_{\mu \bar\sigma} \Gamma^{\bar\sigma}_{\bar\nu \bar\rho} - 2 \partial_\mu W \right) -\frac{1}{2} \bar{G}^{\bar\mu} \left( {\tilde\chi}^\nu {\tilde\chi}^\rho g_{\sigma \bar\mu} \Gamma^{\sigma}_{\nu \rho} - 2 \partial_{\bar\mu} \bar W \right) \nonumber \\ 
&+&\frac{1}{4} \tilde\chi^\mu {\tilde\chi}^\nu \bar{\tilde\chi}^{\bar\rho} \bar{\tilde\chi}^{\bar\sigma} \partial_\nu \partial_{\bar\sigma} g_{\mu \bar\rho} -(\partial_m \psi^\mu \partial^m \bar \psi^{\bar\nu}) g_{\mu \bar\nu} -i \bar{\tilde\chi}^{\bar\mu} \sigma^m g_{\nu \bar\mu} D_m \tilde\chi^\nu \nonumber \\ 
&-&\frac{1}{2} (\partial_\mu \partial_\nu W) \tilde\chi^{\alpha \mu} \tilde\chi^\nu_\alpha -\frac{1}{2} (\partial_{\bar\mu} \partial_{\bar\nu} \bar W) \bar{\tilde\chi}^{\bar\mu}_{\dot\alpha} \bar{\tilde\chi}^{\dot\alpha \bar\nu}.
\end{eqnarray}
The auxiliar fields $G^{\nu}$ and $\bar{G}^{\bar\nu}$ can be integrated 
using the Euler-Lagrange equations, thus we have
\begin{eqnarray}
\label{gec}
\bar{G}^{\bar\nu}= \frac{1}{2} g^{\mu \bar\nu} \left( \bar{\tilde\chi}^{\bar\theta} \bar{\tilde\chi}^{\bar\rho} g_{\mu \bar\sigma} \Gamma^{\bar\sigma}_{\bar\theta \bar\rho} - 2 \partial_\mu W \right), \\
\label{ge}
G^{\nu}= \frac{1}{2} g^{\nu \bar\mu} \left( \tilde\chi^\theta \tilde\chi^\rho g_{\bar\mu \sigma} \Gamma^{\sigma}_{\theta \rho} - 2 \partial_{\bar\mu} \bar{W} \right).
\end{eqnarray}

\section{Examples}

\label{five}

In this Section we present an example taken form \cite{Hori:2000kt} which constitutes the building block to study
Mirror Symmetry in complete intersection Calabi-Yau varieties. This kind of Lagrangian, in particular the twisted superpotential,
also appears when exploring non-Abelian T-dualities in gauged linear sigma models \cite{CaboBizet:2017fzc}.
This is the second example we present, a non-Abelian $SU(2)$ dual model
of the $U(1)$ GLSM with $\mathbb{CP}^1$ target space.

\subsection{Abelian T-dual of a single chiral superfield}
Let us start with the Lagrangian
\begin{eqnarray}
L&=&\int d\theta^4\left(-\frac{1}{2e^2}\bar\Sigma\Sigma-\frac{1}{2}(Y+\bar Y)\ln(Y+\bar Y)\right)\nn \\
&+&\frac{1}{2}\left(\int d^2\tilde\theta \Sigma(qY-t)+c.c.\right)+\frac{1}{2} \mu \left(\int d^2\tilde\theta e^{-Y}+c.c.\right)\label{lagrangianDual}
\end{eqnarray}
given in terms of the twisted chiral superfield $Y$. The twisted superpotential from this expression is $W=\frac{1}{2}(q\Sigma Y- \Sigma t + \mu e^{-Y})$. $q$ denotes the charge of the dual chiral superfield. From the previously developed
expressions we compute using equation $(\ref{metric})$ the K\"ahler metric and its inverse

\begin{eqnarray}
g_{\mu \bar{\nu}}&=& \left[ \begin{array}{cc} 
\frac{1}{2e^2} & 0\\
0 & \frac{1}{2(y + \bar{y})}\\
\end{array}
\right], \\
g^{\bar{\mu} \nu}&=& \left[ \begin{array}{cc} 
2e^2 & 0\\
0 & 2(y + \bar{y})\\
\end{array}
\right] ,
\end{eqnarray}
where $\mu,\nu, \ldots$ runs over $1,2$ while $\bar\mu,\bar\nu, \ldots$ over $\bar{1},\bar{2}$. The index $1$ and $2$ are identified with the fields $\sigma$ (scalar component of $\Sigma$) and $y$ (scalar component of $Y$) respectively, the barred indices are assocciated
to their complex conjugates. The Cristoffel symbols can be computed using the expressions $(\ref{chris})$ and $(\ref{chrisc})$. The only non vanishing terms are

\begin{equation}
\Gamma^{2}_{2 2}=\Gamma^{\bar2}_{\bar2 \bar2}=-\frac{1}{y+\bar{y}}.
\end{equation}
The auxiliary fields can be easily computed from expressions $(\ref{gec})$ and $(\ref{ge})$

\begin{eqnarray}
\bar{G}^{\bar1}=e^2(-q y+ t), \quad \bar{G}^{\bar2}=-\frac{1}{2(y+\bar{y})} \bar{\tilde\chi}^{\bar 2}_{\dot\alpha} \bar{\tilde\chi}^{\bar 2 \dot\alpha} +(\mu e^{-y}-q\sigma)(y +\bar y),\\
G^{1}=e^2(-q \bar{y}+ \bar{t}), \quad G^{2}=-\frac{1}{2(y+\bar{y})} \tilde\chi^{2 \dot\alpha} \tilde\chi^{2}_{\alpha} +(\mu e^{-\bar{y}} -q \bar{\sigma})(y +\bar y ).
\end{eqnarray}
Thus the contributing terms to the Lagrangian $(\ref{lagrangianDual})$ are

\begin{eqnarray}
\mathcal{L}&=&G^1 \bar G^1 g_{1 \bar 1}+G^2 \bar G^{\bar2} g_{2 \bar2} -\frac{1}{2} G^1 \left( - 2 \partial_1 W \right) - \frac{1}{2} \bar G^{\bar1} \left( - 2 \partial_{\bar1} \bar W \right) \\ &-&\frac{1}{2} G^2 \left( \bar{\tilde\chi}^{\bar2} \bar{\tilde\chi}^{\bar2} g_{2 \bar2} \Gamma^{\bar2}_{\bar2 \bar2} - 2 \partial_2 W \right)-\frac{1}{2} \bar{G}^{\bar2} \left( {\tilde\chi}^2 {\tilde\chi}^2 g_{2 \bar2} \Gamma^{2}_{2 2} - 2 \partial_{\bar2} \bar W \right) \nonumber \\ 
& -& (\partial_m \sigma \partial^m \bar \sigma) g_{1\bar1} -(\partial_m y \partial^m \bar y) g_{2 \bar2}-i \bar{\tilde\chi}^{\bar 1} \sigma^m g_{1 \bar1} D_m \tilde\chi^1 -i \bar{\tilde\chi}^{\bar 2} \sigma^m g_{2 \bar2} D_m \tilde\chi^2 \nn \\
&-&\frac{1}{2} (\partial_2 \partial_2 W) \tilde\chi^{\alpha 2} \tilde\chi^2_{\alpha} - \frac{1}{2} (\partial_{\bar2} \partial_{\bar2} \bar W) \bar{\tilde\chi}^{\bar2}_{\dot\alpha} \bar{\tilde\chi}^{\dot\alpha \bar2}+\frac{1}{4} \tilde\chi^2 {\tilde\chi}^2 \bar{\tilde\chi}^{\bar2} \bar{\tilde\chi}^{\bar2} \partial_2 \partial_{\bar 2} g_{2 \bar2} . \nn
\end{eqnarray}
Substituting the metric component, the Christoffel symbols and integrating the auxiliary fields the Lagrangian is
written as
\begin{eqnarray}
&&\mathcal{L}= \frac{1}{2}e^2(-q \bar{y}+ \bar{t})(-q y+ t) \nonumber \\ 
&+& \frac{(y+\bar{y})}{2} \left( \frac{-\bar{\tilde\chi}^{\bar2}_{\dot\alpha} \bar{\tilde\chi}^{\bar2 \dot\alpha} }{2(y+\bar y)^2}  + \mu e^{-y}-q \sigma \right) \left( \frac{-\tilde\chi^{2 \alpha} \tilde\chi^{2}_{\alpha}}{2(y+\bar y)^2}  + \mu e^{-\bar y}-q \bar\sigma \right) \nonumber \\ 
&-&\frac{1}{2}e^2(-q \bar{y}+ \bar{t})(-q y+ t)-\frac{1}{2}e^2(-q y+ t)(-q \bar{y}+ \bar{t}) \nonumber \\ 
&-&\frac{(y +\bar{y})}{2} \left( \frac{- \tilde\chi^{2 \alpha} \tilde\chi^{2}_{\alpha}}{2(y+\bar{y})^2} +\mu e^{-\bar{y}}-q \bar \sigma \right) \left( \frac{-\bar{\tilde\chi}^{\bar2}_{\dot\alpha} \bar{\tilde\chi}^{\bar2  \dot\alpha}}{2(y+ \bar y)^2}  + \mu e^{-y}-q\sigma \right) \nonumber \\ 
&-&\frac{(y +\bar{y})}{2} \left( \frac{- \bar{\tilde\chi}^{\bar2}_{\dot\alpha} \bar{\tilde\chi}^{\bar2  \dot\alpha}}{2(y+\bar y)^2}  + \mu e^{-y}-q \sigma \right) \left( \frac{- {\tilde\chi}^{2 \alpha} {\tilde\chi}^2_{\alpha}}{2(y+ \bar y)^2} + \mu e^{-\bar y}-q\bar{\sigma} \right) \nonumber \\ 
&+&\frac{1}{4} \tilde\chi^{2 \alpha} {\tilde\chi}^2_{\alpha} \bar{\tilde\chi}^{\bar2}_{\dot\alpha} \bar{\tilde\chi}^{\bar2 \dot\alpha} \frac{1}{(y+ \bar y)^3} -\frac{1}{2 e^2}(\partial_m \sigma \partial^m \bar \sigma) -\frac{1}{2(y+ \bar y)}(\partial_m y \partial^m \bar y) \nonumber \\
&-& i \frac{1}{2 e^2}\bar{\tilde\chi}^{\bar 1 \dot\alpha} \sigma^m_{\alpha \dot\alpha} D_m \tilde\chi^{1\alpha}
-i \frac{1}{2(y+ \bar y)}\bar{\tilde\chi}^{\bar 2 \dot\alpha} \sigma^m_{\alpha \dot\alpha} D_m \tilde\chi^{2 \alpha} -\frac{\mu e^{-y}}{4} \tilde\chi^{2 \alpha} \tilde\chi^2_{\alpha} \nonumber \\ 
&-& \frac{\mu e^{-\bar y}}{4} \bar{\tilde\chi}^{\bar2}_{\dot\alpha} \bar{\tilde\chi}^{\bar2 \dot\alpha}.
\end{eqnarray}
Recall that the symbols $\partial_m$ represent space-time derivatives. After simplifying, the final Lagrangian for this model is given by

\begin{eqnarray}
&&\mathcal{L}=-\frac{e^2}{2} ||-qy+t ||^2+\frac{1}{8(y+ \bar{y})^3} \tilde\chi^{2 \alpha} {\tilde\chi}^2_{\alpha} \bar{\tilde\chi}^{\bar2}_{\dot\alpha} \bar{\tilde\chi}^{\bar2 \dot\alpha} -\frac{1}{2 e^2}(\partial_m \sigma \partial^m \bar \sigma) \nonumber \\ &-& \frac{1}{2(y+ \bar y)}(\partial_m y \partial^m \bar y) -i \frac{1}{2 e^2}\bar{\tilde\chi}^{\bar 1 \dot\alpha} \sigma^m_{\alpha \dot\alpha} D_m \tilde\chi^{1 \alpha} -i \frac{1}{2(y+ \bar y)}\bar{\tilde\chi}^{\bar 2 \dot\alpha} \sigma^m_{\alpha \dot\alpha} D_m \tilde\chi^{2 \alpha} \nonumber \\&-& \frac{\mu e^{-y}}{4} \tilde\chi^{\alpha 2} \tilde\chi^2_{\alpha} -\frac{\mu e^{- \bar y}}{4} \bar{\tilde\chi}^{\bar2}_{\dot\alpha} \bar{\tilde\chi}^{\dot\alpha \bar2} +\frac{(-q \sigma+\mu e^{-y})}{4(y+\bar y)} \tilde\chi^{2 \alpha} \tilde\chi^{2}_{\alpha} + \frac{(-q \bar \sigma+\mu e^{-\bar y})}{4(y+\bar y)} \bar{\tilde\chi}^{\bar2}_{\dot\alpha} \bar{\tilde\chi}^{\bar2 \dot\alpha} \nonumber \\ &-& \frac{1}{2} (q^2 ||\sigma||^2+ \mu^2
e^{-(y+\bar y)}) (y+\bar y)+\frac{\mu q}{2}(\sigma e^{-\bar{y}}+\bar\sigma e^{-y}) (y+\bar{y}).
\end{eqnarray}

This completes the Abelian-dual Lagrangian to a single chiral superfield $U(1)$ GLSM in \cite{Hori:2000kt}, including the fermionic contributions. Recall
the fermionic components are given in the twisted notation of Section \ref{three}.
This action constitutes the building block of the physics proof of Mirror Symmetry \cite{Hori:2000kt}
for complete intersection Calabi-Yau varieties.

\subsection{Non-Abelian $SU(2)$ dual of $\mathbb{CP}^1$}

In this Subsection we present the calculation of the Lagrangian in components for a non-Abelian SU(2) dual obtained
in \cite{CaboBizet:2017fzc}. This is the building block to dualize the deformed conifold GLSM under the non-Abelian $SU(2)\times SU(2)$ global symmetry. The resulting model is a non-Abelian T-dual. The bosonic terms were calculated in \cite{CaboBizet:2017fzc}. Here we write both the
bosonic and fermionic contributions to the Lagrangian.
We start with the dual Lagrangian 
\begin{eqnarray}
\mathcal{L}_{dual}&=&\int d^4\theta \sqrt{( Y^1 + \bar Y^1)^2+ (Y^2+\bar Y^2)^2+ (Y^3 + \bar Y^3)^2}+\label{exT}\\
&+&\left( \int d^4\theta Y^\mu \hat n_\mu \ln(\mathcal{K}(Y^i,\bar Y^i,n_j))+c.c.\right)- \int d^4\theta\frac{1}{2 e^2}\bar{\Sigma}^0\Sigma^0\nn\\
&+&q\int d\bar \theta^-d\theta^+\left(Y^\mu n_\mu-\frac{t}{2 q}\right) \Sigma^0 +c.c.\nn\\
&+&\frac{q}{2}\int d\bar \theta^-d\theta^+\left(Y^\mu \bar D_+n_\mu \right) D_-V^0+c.c.\nn
\end{eqnarray}
This depends on the twisted (anti-)chiral superfields $Y^\mu$($\bar Y^\mu$) with $\mu=1,2,3$, the
twisted field strength $\Sigma^0$ and on semi-chiral real superfields $n_\mu$ \cite{Bogaerts:1999jc,Gates:2007ve}, which for our example calculation we consider to be constant. \footnote{The component $D_- V_0$ is not a twisted chiral superfield, so there will be additional
contributions to the ones computed here.}
The twisted chiral superfield has an expansion

\begin{eqnarray}
\label{XaTwist}
Y^\mu&=& y^\mu(x)+ \sqrt{2} \tilde\theta^{\alpha} \chi^\mu_{\alpha}(x)+\tilde\theta^2 G^\mu(x)+\frac{i}{\sqrt{2}} \tilde\theta^2 \bar{\tilde\theta}^{\dot\alpha} \sigma^m_{\alpha \dot\alpha} \partial_m \tilde\chi^{\mu \alpha} \nonumber \\
&&i \tilde\theta^\alpha \sigma^m_{\alpha \dot\alpha} \bar{\tilde\theta}^{\dot\alpha}\partial_m y^\mu(x) -\frac{1}{4}\tilde\theta^2 \bar{\tilde\theta}^2 \square y^\mu (x).
\end{eqnarray}
The field strength of the $U(1)$ vector superfield $V^0$ is a twisted chiral superfield and its expansion 
can be written in a similar fashion as:
\begin{eqnarray}
\label{sigma0}
\Sigma^0&=& \sigma^0(x)+ \sqrt{2} \tilde\theta^{\alpha} \chi^0_{\alpha}(x)+\tilde\theta^2 G^0(x)+\frac{i}{\sqrt{2}} \tilde\theta^2 \bar{\tilde\theta}^{\dot\alpha} \sigma^m_{\alpha \dot\alpha} \partial_m \tilde\chi^{0\alpha(x)} \nonumber \\
&&i \tilde\theta^\alpha \sigma^m_{\alpha \dot\alpha} \bar{\tilde\theta}^{\dot\alpha}\partial_m \sigma^0(x) -\frac{1}{4}\tilde\theta^2 \bar{\tilde\theta}^2 \square \sigma^0 (x).
\end{eqnarray}
In the standard notation for the two dimensional field strength the auxiliary field is given
in terms of the $D$ term and the Yang-Mills field strength $v_{03}$ by 
the expression $G^0=D-i v_{03}$.

The metric and the connection in K\"ahler field space are given by:
\begin{eqnarray}
g_{\mu \bar \nu}&=&\frac{\partial}{\partial y^\mu \bar y^{\bar \nu}}K. \\
&=&\frac{(y^\mu+\bar y^\mu)(y^\nu+\bar y^\nu)}{\sqrt{\sum_\mu(y^\mu+\bar y^\mu)^2}} +n_\mu\partial_{\bar{\nu}} \ln\mathcal{K}+n_\nu \partial_{\mu} \ln\mathcal{\bar K}\nn\\
&+&y^\rho n_\rho \frac{\partial_\mu \partial_{\bar \nu}\mathcal{K}}{\mathcal{K}}+\bar y^\rho n_\rho \frac{\partial_\mu \partial_{\bar \nu} \bar{\mathcal{K}}}{\bar{\mathcal{K}}}-\bar y^\rho n_\rho \frac{\partial_\mu \bar{\mathcal{K}}\partial_{\bar \nu}\bar{\mathcal{K}}}{\bar{\mathcal{K}}^2}
- y^\rho n_\rho \frac{\partial_\mu{\mathcal{K}}\partial_{\bar \nu}{\mathcal{K}}}{{\mathcal{K}}^2}\nn\\
&+& \delta_{\mu,\nu}\frac{1}{\sqrt{\sum_\mu(y^\mu+\bar y^\mu)^2}},\nn\\
\Gamma^\nu_{\mu\rho}&=&g^{\nu \bar \mu}\frac{\partial}{\partial y^\rho} g_{\mu \bar \mu},\nonumber\\
g_{0\bar 0}&=&-1/(2 e^2), \, \,  \,   \,  \, \Gamma^0_{00}=0.\nn
\end{eqnarray}

This leads to the Lagrangian for fermionic and bosonic components
\begin{eqnarray}
\label{lag_f2}
\mathcal{L}&=&G^\mu \bar G^{\bar \nu} g_{\mu \bar \nu} +G^0 \bar G^{\bar 0} g_{0 \bar 0} \\
&-&\frac{1}{2} G^\mu \left( \bar{\tilde\chi}^{\bar \nu} \bar{\tilde\chi}^{\bar \rho} g_{\mu \bar\sigma} \Gamma^{\bar\sigma}_{\bar \nu \bar \mu} - 2 q n_\mu \sigma^0 \right) -\frac{1}{2} \bar{G}^{\bar \mu} \left( {\tilde\chi}^\nu {\tilde\chi}^\rho g_{\sigma \bar \mu} \Gamma^{\sigma}_{\nu \rho} - 2 q n_\mu \bar \sigma^0 \right) \nonumber \\ &+&\frac{1}{4} \tilde\chi^\mu {\tilde\chi}^\nu \bar{\tilde\chi}^{\bar \rho} \bar{\tilde\chi}^{\bar \sigma} \partial_\nu \partial_{\bar \sigma} g_{\mu \bar \rho} -(\partial_m y^\mu \partial^m \bar y^{\bar \nu}) g_{\mu \bar \nu} +(\partial_m \sigma^0 \partial^m \bar \sigma ^{\bar 0}) /(2 e^2)\nn\\
& -&i \bar{\tilde\chi}^{\bar \mu} \sigma^m g_{\nu \bar \mu} D_m \tilde\chi^{\nu}
+i \bar{\tilde\chi}^{\bar 0} \sigma^m D_m \tilde\chi^0/(2 e^2) \nonumber \\ &-&n_\mu q \tilde\chi^{\mu \alpha } \tilde\chi^0_\alpha -n_\mu q \bar{\tilde\chi}^{\bar \mu}_{\dot\alpha} \bar{\tilde \chi}^{\dot\alpha \bar 0}\nn \\
&+&qG^0 (y^\mu n_\mu-t/(2q))+q\bar G^0 (\bar y^\mu  n_\mu-t/(2q)).\nn
\end{eqnarray}
Recall that fermions are given in our twisted notation; for example the components express as $\tilde\chi_{\alpha}^\mu=(\bar \chi_{+}^\mu,\chi_{-}^\mu)$
and $\tilde\chi^{\mu \alpha}=(\chi_{-}^\mu, -\bar\chi_{+}^\mu)$ with $\alpha=+,-$.  Also notice that the coefficients $n_{\mu}$ are real,
so no complex conjugate of them appears. This matches the non kinetic components obtained for this model in \cite{CaboBizet:2017fzc}.
Auxiliary fields are still not integrated. It is the matter of future work to employ these components in order to study the $SU(2)$ non-Abelian 
dual model partition functions on the sphere. For this the fermionic components are crucial.



\section{Actions for chiral and twisted chiral}
\label{six}

In this Section we compute the Lagrangian of a theory with mixed kinetic terms of chiral and twisted chiral superfields. These are
generic Lagrangians in 2D $(2,2)$ supersymmetric theories. Calculations are involved, so we cite the main result
and relegate the formulas including fermions to the Appendix \ref{two}.

Let us write the chiral superfields expansion \cite{wessbagger}:
\begin{eqnarray}
\Phi (x)&=& \phi(x)+i \theta^\alpha \sigma^m_{\alpha \dot\alpha} \bar{\theta}^{\dot\alpha}\partial_m \phi(x) -\frac{1}{2} \theta^\alpha \sigma^m_{\alpha \dot\alpha} \bar{\theta}^{\dot\alpha} \theta^\beta \sigma^n_{\beta \dot\beta} \bar{\theta}^{\dot\beta} \partial_m \partial_n \phi(x) +\nonumber \\ 
&&\sqrt{2} \theta^{\alpha}[\chi_{\alpha}(x)+i \theta^\beta \sigma^m_{\beta \dot\beta} \bar{\theta}^{\dot\beta} \partial_m \chi_\alpha(x)] + \theta^{\alpha} \theta_{\alpha} F(x).
\end{eqnarray}
Notice that in this Section, $\chi_{\alpha}$ are the fermionic components for chiral superfields, in contrast we denote by ${\tilde\chi}_{\alpha}$ the fermionic components for twisted chiral superfields in (\ref{twsf}). The anti-chiral superfield expansion is given by:

\begin{eqnarray}
\bar\Phi (x)&=& \bar\phi(x)-i \theta^\alpha \sigma^m_{\alpha \dot\alpha} \bar{\theta}^{\dot\alpha}\partial_m \bar\phi(x) -\frac{1}{2} \theta^\alpha \sigma^m_{\alpha \dot\alpha} \bar{\theta}^{\dot\alpha} \theta^\beta \sigma^n_{\beta \dot\beta} \bar{\theta}^{\dot\beta} \partial_m \partial_n \bar\phi(x) +\nonumber \\ 
&&\sqrt{2} \bar\theta_{\dot\alpha}[\bar\chi^{\dot\alpha}(x)-i \theta^\beta \sigma^m_{\beta \dot\beta} \bar{\theta}^{\beta} \partial_m {\bar\chi}^{\dot\alpha}(x)] + \bar\theta_{\dot\alpha} \bar\theta^{\dot\alpha} \bar F(x).
\end{eqnarray}

A generic term in the K\"ahler potential can be written as a sum of powers of chiral $\Phi^\rho$, anti-chiral $\bar \Phi^\eta$, twisted chiral $\Psi^\mu$ and twisted anti-chiral $\bar\Psi^\nu$ superfields. This yields a series expansion as follows:

\begin{eqnarray}
\label{kexpanCT}
K(\Psi^\mu,\bar\Psi^\nu,\Phi^\rho,\bar\Phi^\eta)&=&\sum_{i,j,m,n}\sum_{\underline{\mu},\underline{\nu},\underline{\rho},\underline{\eta}} c_{\mu_1\cdots \mu_i \nu_1\cdots\nu_j \rho_1\cdots \rho_m \eta_1\cdots\eta_n} \Psi^{\mu_1} \cdots \Psi^{\mu_i} \nn \\
&&\bar\Psi^{\nu_1}\cdots \bar\Psi^{\nu_j} \times \Phi^{\rho_1} \cdots \Phi^{\rho_m}\bar\Phi^{\eta_1} \cdots \bar\Phi^{\eta_n},\label{kaehlerTot} \\
L_{tot}&=&\int d^4\theta K(\Psi^\mu,\bar\Psi^\nu,\Phi^\rho,\bar\Phi^\eta).\label{LTot}
\end{eqnarray}
The indices $i,j,m,n$ denote the number of chiral, anti-chiral, twisted-chiral and twisted anti-chiral superfields. 
The indices $\mu_k, \nu_k,\rho_k,\eta_k$ label a given superfield and run from 1 to the total number of the respective superfield. 
The notation $\underline{\mu}$ means the tuple $\mu_1,...,\mu_i$ etc.
A generic function for chiral superfields can be expressed as:
\begin{equation}
\label{polyC1}
P (\Phi)= P(\phi) + \sqrt{2} {\theta}^{ \alpha} {\chi}^{\mu}_{\alpha} \frac{\partial P}{\partial \bar \phi^{\mu} } + {\theta}^2 \left[ F^\mu \frac{\partial P}{\partial \phi^{\mu} } -\frac{1}{2} \frac{\partial^2 P}{\partial \phi^\mu \partial \phi^\nu} {\chi}^{\mu}_{\alpha} {\chi}^{\nu \alpha } \right].
\end{equation}
This building block has the same structure as the superpotential. Also its conjugate can be written as:
\begin{equation}
\label{polyC2}
\bar P (\bar\Phi)= \bar P(\bar \phi) + \sqrt{2} {\bar\theta}^{\dot \alpha} \bar{\chi}^{\bar\mu}_{\dot \alpha} \frac{\partial \bar P}{\partial \bar \phi^{\bar\mu} } + \bar{\theta}^2 \left[ \bar F^{\bar\mu} \frac{\partial \bar P}{\partial \bar \phi^{\bar\mu} } -\frac{1}{2} \frac{\partial^2 \bar P}{\partial \bar \phi^{\bar\mu} \partial \bar \phi^{\bar\nu}} \bar{\chi}^{\bar\mu}_{\dot\alpha} \bar{\chi}^{\bar\nu \dot\alpha} \right].
\end{equation}

Let us consider the Lagrangian:
\begin{eqnarray}
L^{\underline{\mu},\underline{\nu},\underline{\rho},\underline{\eta}}&=&\int d\theta^4\Phi^{\mu_1}\cdots \Phi^{\mu_i}\bar\Phi^{\nu_1}\cdots \bar\Phi^{\nu_j}\Psi^{\rho_1}\cdots \Psi^{\rho_m} \bar\Psi^{\eta_1}\cdots \bar\Psi^{\eta_n},\label{mixed} \\
&=& \int d\theta^4 \kappa^{\underline{\mu},\underline{\nu},\underline{\rho},\underline{\eta}}. \nn
\end{eqnarray}
which is a single summand of (\ref{kexpanCT}). The sum over all
terms in the series (\ref{kaehlerTot}) yields the most generic Lagrangian i.e.  $L_{tot}=
\sum_{i,j,m,n}\sum_{\underline{\mu},\underline{\nu},\underline{\rho},\underline{\eta}} c_{\underline{\mu},\underline{\nu},\underline{\rho},\underline{\eta}} L^{\underline{\mu},\underline{\nu},\underline{\rho},\underline{\eta}}$. The extra ingredient
is that an arbitrary combination  of chiral and twisted chiral superfields can be treated. The other terms in the Lagrangian are in the superpotential (depending only on chiral superfields) or in the twisted superpotential (depending only
on twisted chiral superfields). Now we proceed to realize a Taylor expansion in the Grassman variables

\begin{eqnarray}
\kappa^{\underline{\mu},\underline{\nu},\underline{\rho},\underline{\eta}}=\Phi^{\mu_1}\cdots \Phi^{\mu_i}\bar\Phi^{\nu_1}\cdots \bar\Phi^{\nu_j}\Psi^{\rho_1}\cdots \Psi^{\rho_m} \bar\Psi^{\eta_1}\cdots \bar\Psi^{\eta_n}= \label{ldens} \\
\left( P + \sqrt{2} \theta^{\alpha} \chi_{\alpha}^{\mu} \partial_\mu P +\theta^{\alpha} \theta_{\alpha} \left[ F^\mu \partial_\mu P -\frac{1}{2}\chi^{\mu} \chi^{\nu} \partial_\mu \partial_\nu P \right] \right) \cdot \nonumber \\ 
\left( \bar{P} + \sqrt{2} \bar{\theta}_{\dot\alpha} \bar{\chi}^{\bar\nu \dot\alpha} \partial_{\bar\nu} \bar P +\bar{\theta}_{\dot\alpha} \bar{\theta}^{\dot\alpha} \left[ \bar{F}^{\bar\mu} \partial_{\bar\mu} \bar{P} -\frac{1}{2}\bar\chi^{\bar\mu} \bar\chi^{\bar\nu} \partial_{\bar\mu} \partial_{\bar\nu} \bar{P} \right] \right) \cdot \nonumber \\
\left( Q + \sqrt{2} \tilde\theta^{\alpha} \tilde\chi_{\alpha}^{\rho} \partial_\rho Q +\tilde\theta^{\alpha} \tilde\theta_{\alpha} \left[ G^\mu \partial_\mu Q -\frac{1}{2}\tilde\chi^{\mu} \tilde\chi^{\nu} \partial_\mu \partial_\nu Q \right] \right) \cdot \nonumber \\ 
\left( \bar{Q} + \sqrt{2} \bar{\tilde\theta}_{\dot\alpha} \bar{\tilde\chi}^{\bar\eta \dot\alpha} \partial_{\bar\eta} \bar{Q} +\bar{\tilde\theta}_{\dot\alpha} \bar{\tilde\theta}^{\dot\alpha} \left[ \bar{G}^{\bar\mu} \partial_{\bar\mu} \bar{Q} -\frac{1}{2}\bar{\tilde\chi}^{\bar\mu} \bar{\tilde\chi}^{\bar\nu} \partial_{\bar\mu} \partial_{\bar\nu} \bar{Q} \right] \right).\nn
\end{eqnarray}
In the previous formula $P=\phi^{\mu_1}\cdots \phi^{\mu_i}$, $\bar{P}=\bar\phi^{\nu_1}\cdots \bar\phi^{\nu_j}$,$Q=\psi^{\rho_1}\cdots \psi^{\rho_m}$ and $\bar{Q}=\bar\psi^{\eta_1}\cdots \bar\psi^{\eta_n}$. Notice that $P$ and $Q$ are not the conjugates of $\bar P$ and
$\bar Q$. The polynomials are evaluated in the corresponding scalar componentes of the $\Phi$'s and $\Psi$'s and their complex conjugates. Notice that the greek indices denoting derivarives w.r.t scalars  run over the number of
chiral, anti-chiral, twisted chiral, twisted-antichiral fields when is applied on $P$, $\bar P$, $Q$ and $\bar Q$.
Next, we define the following scalar variables to simplify the computations 

\begin{eqnarray}
f= F^\mu \partial_\mu P -\frac{1}{2}\chi^{\mu} \chi^{\nu} \partial_\mu \partial_\nu P, \, \, \bar{f}= \bar{F}^{\bar\mu} \partial_{\bar\mu} \bar{P} -\frac{1}{2}\bar\chi^{\bar\mu} \bar\chi^{\bar\nu} \partial_{\bar\mu} \partial_{\bar\nu} \bar{P}, \\
g= G^\mu \partial_\mu Q -\frac{1}{2}\tilde\chi^{\mu} \tilde\chi^{\nu} \partial_\mu \partial_\nu Q, \, \, \bar{g}= \bar{G}^{\bar\mu} \partial_{\bar\mu} \bar{Q} -\frac{1}{2}\bar{\tilde\chi}^{\bar\mu} \bar{\tilde\chi}^{\bar\nu} \partial_{\bar\mu} \partial_{\bar\nu} \bar{Q}. \nn
\end{eqnarray}
Thus the argument in the integral (\ref{ldens}), which gives the Lagrangian, can be written as
\begin{eqnarray}
&&\kappa^{\underline{\mu},\underline{\nu},\underline{\rho},\underline{\eta}}=  \label{l1}\\
&&P\bar{P}Q \bar{Q}+ \tilde\theta^{\alpha} \tilde\theta_{\alpha} \bar{\tilde\theta}_{\dot\alpha} \bar{\tilde\theta}^{\dot\alpha} g \bar{g} P \bar{P} +\theta^{\alpha} \theta_{\alpha} \bar\theta_{\dot\alpha} \bar\theta^{\dot\alpha} f \bar{f} Q \bar{Q} \nn\\
&&+ 2 \theta^{\alpha} \chi^{\mu}_{\alpha} \left(\partial_{\mu} P \right) \bar\theta_{\dot\alpha} \bar\chi^{\bar\nu \dot\alpha} \left(\partial_{\bar\nu} \bar{P} \right) Q \bar{Q} + 2 \tilde\theta^{\alpha} \tilde\chi^{\mu}_{\alpha} \left( \partial_\mu Q\right) \bar{\tilde\theta}_{\dot\alpha} \bar{\tilde\chi}^{\bar\nu \dot\alpha} \left( \partial_{\bar\nu} \bar{Q} \right) P \bar{P} \nn \\ 
&&+ 2 \bar\theta_{\dot\alpha} \bar\chi^{\bar\mu \dot\alpha} \left( \partial_{\bar\mu} \bar{P}\right) \bar{\tilde\theta}_{\dot\beta} \bar{\tilde\chi}^{\bar\nu \dot\beta} \left(\partial_{\bar\nu} \bar{Q} \right) P Q+2 \theta^{\alpha} \chi^{\mu}_{\alpha} \left(\partial_{\mu} P \right) \tilde\theta^{\beta} \tilde\chi^{\nu}_{\beta} \left( \partial_\nu Q\right) \bar{P} \bar{Q}  \nonumber \\
&&+ 2 \theta^{\alpha} \chi^{\mu}_{ \alpha} \left(\partial_{\mu} P \right) \bar{\tilde\theta}_{\dot\alpha} \bar{\tilde\chi}^{\bar\nu \dot\alpha} \left( \partial_{\bar\nu} \bar{Q}\right) \bar{P} Q + 2 \bar\theta_{\dot\alpha} \bar\chi^{\bar\mu \dot\alpha} \left( \partial_{\bar\mu} \bar{P} \right) \tilde\theta^{\alpha} \tilde\chi^{\nu}_{\alpha} \left( \partial_\nu Q \right) P \bar{Q}\nonumber\\
&&+2 \bar{\tilde\theta}_{\dot\alpha} \bar{\tilde\theta}^{\dot\alpha} \theta^{\alpha} \chi^{\mu}_{ \alpha} \left(\partial_{\mu} P \right) \bar\theta_{\dot\beta} \bar\chi^{\bar\nu \dot\beta} \left( \partial_{\bar\nu} \bar{P}\right) \bar{g} Q+2 \tilde\theta^{\alpha} \tilde\theta_{\alpha} \theta^{\beta} \chi^{\mu}_{\beta} \left(\partial_{\mu} P \right) \bar\theta_{\dot\beta} \bar\chi^{\bar\nu \dot\beta} \left( \partial_{\bar\nu} \bar{P}\right)g \bar{Q} \nonumber \\
&&+ 2 \theta^{\alpha} \theta_{\alpha} \tilde\theta^{\beta} \tilde\chi^{\mu}_{\beta}(\partial_{\mu} Q)\bar{\tilde\theta}^{\dot{\alpha}} \bar{\tilde\chi}^{\bar{\nu}} (\partial_{\bar{\nu}} \bar{Q})f \bar{P} + 2 \bar\theta_{\dot\alpha} \bar\theta^{\dot\alpha} \tilde\theta^{\alpha} \tilde\chi^{\mu}_{ \alpha} \left(\partial_{\mu} Q \right) \bar{\tilde\theta}_{\dot\alpha} \bar{\tilde\chi}^{\bar\nu \dot\alpha} \left( \partial_{\bar\nu} \bar{Q}\right) P \bar{f} \nonumber\\
&&+ 2 \bar{\tilde\theta}_{\dot\alpha} \bar{\tilde\theta}^{\dot\alpha} \theta^{\alpha} \chi^{\mu}_{ \alpha} \left(\partial_{\mu} P \right) \tilde\theta^{\beta} \tilde\chi^{\nu}_{\beta} \left( \partial_\nu Q\right) \bar{P} \bar{g}+2 \tilde\theta^{\alpha} \tilde\theta_{\alpha} \bar\theta_{\dot\beta} \bar\chi^{\bar\mu \dot\beta} \left(\partial_{\bar\mu} \bar{P} \right) \bar{\tilde\theta}_{\dot\alpha} \bar{\tilde\chi}^{\bar\nu \dot\alpha} \left( \partial_{\bar\nu} \bar{Q}\right) P g \nonumber \\
&&+ 2 \theta^{\alpha} \theta_{\alpha} \bar\theta_{\dot\beta} \bar\chi^{\bar\mu \dot\beta} \left(\partial_{\bar\mu}\bar{P} \right) \tilde\theta^{\beta} \tilde\chi^{\nu}_{\beta} \left( \partial_\nu Q\right) \bar{Q} f + 2 \bar\theta_{\dot\alpha} \bar\theta^{\dot\alpha} \theta^{\beta} \chi^{\mu}_{ \beta} \left(\partial_{\mu} P \right) \bar{\tilde\theta}_{\dot\beta} \bar{\tilde\chi}^{\bar\nu \dot\beta} \left( \partial_{\bar\nu} \bar{Q}\right) Q \bar{f}\nonumber \\
&&+ 2 \bar\theta_{\dot\alpha} \bar\theta^{\dot\alpha} \theta^{\alpha} \chi^{\mu}_{ \alpha} \left(\partial_{\mu} P \right) \tilde\theta^{\beta} \tilde\chi^{\nu}_{\beta} \left( \partial_\nu Q \right) \bar{Q} \bar{f} + 2 \theta^{\alpha} \theta_{\alpha} \bar\theta_{\dot\beta} \bar\chi^{\bar\mu \dot\beta} \left(\partial_{\bar\mu} \bar{P} \right) \bar{\tilde\theta}_{\dot\beta} \bar{\tilde\chi}^{\bar\nu \dot\beta} \left( \partial_{\bar\nu} \bar{Q}\right) Q f\nonumber \\
&&+ 2 \bar{\tilde\theta}_{\dot\alpha} \bar{\tilde\theta}^{\dot\alpha} \bar\theta_{\dot\beta} \bar\chi^{\bar\mu \dot\beta} \left(\partial_{\bar\mu} \bar{P} \right) \tilde\theta^{\alpha} \tilde\chi^{\nu}_{\alpha} \left( \partial_\nu Q\right) P \bar{g}+   2 \tilde\theta^{\alpha} \tilde\theta_{\alpha} \theta^{\beta} \chi^{\mu}_{ \beta} \left(\partial_{\mu} P \right) \bar{\tilde\theta}_{\dot\alpha} \bar{\tilde\chi}^{\bar\nu \dot\alpha} \left( \partial_{\bar\nu} \bar{Q}\right) \bar{P} g \nn \\
&&+4 \theta^{\alpha} \chi^{\mu}_{\alpha} \left( \partial_{\mu} P \right) \bar\theta_{\dot\alpha} \bar\chi^{\bar\nu \dot\alpha} \left( \partial_{\bar\nu} \bar{P} \right) \tilde\theta^{\beta} \tilde\chi^{\rho}_{\beta} \left( \partial_{\rho} Q \right) \bar{\tilde\theta}_{\dot\beta} \bar{\tilde\chi}^{\bar\eta \dot\beta} \left( \partial_{\bar\eta} \bar{Q} \right) \nonumber \\
&&+\theta^{\alpha} \theta_{\alpha} f\bar{P} Q\bar{Q} +\bar\theta_{\dot\alpha} \bar\theta^{\dot\alpha} \bar{f} P Q\bar{Q}+
\tilde\theta^{\alpha} \tilde\theta_{\alpha} P \bar{P} \bar{Q} g + \bar{\tilde\theta}_{\dot\alpha} \bar{\tilde\theta}^{\alpha} P \bar{P} Q \bar{g} .\nn \\
&&+ \theta^{\alpha} \theta_{\alpha} \bar{\tilde\theta}_{\dot\alpha} \bar{\tilde\theta}^{\dot\alpha} f \bar{g} Q \bar{P} + \theta^{\alpha} \theta_{\alpha} \tilde\theta^{\beta} \tilde\theta_{\beta} f g \bar{Q} \bar{P} \nn\\
&&+ + \bar\theta_{\dot\alpha} \bar\theta^{\dot\alpha} \bar{\tilde\theta}_{\dot\beta} \bar{\tilde\theta}^{\dot\beta} \bar{f} \bar{g} P Q + \bar\theta_{\dot\alpha} \bar\theta^{\dot\alpha} \tilde\theta^{\alpha} \tilde\theta_{\alpha} \bar{f} g P \bar{Q}. \nn
\end{eqnarray}
Recall the variables on which the superfields depend are the redefined space-time coordinates $X^{m},\bar X^{\bar m}$ (\ref{st-tq}) and $\tilde X^{m},\bar{\tilde{X}}^{\bar m}$ (\ref{coord1}). On previous expansion we consider only the terms proportional to $\theta^0$, $\theta^2$ and $\theta^4$, which are the only
ones that contribute to the K\"ahler potential. Now we proceed to expand the polynomials $P, \bar{P}, Q$ and $\bar{Q}$ in terms of the space-time variables using the expressions $(\ref{st-tq})$. Let us take into account that using the chain rule the redefinitions of the kind:
$P \bar{P} Q (\partial_m \partial^m \bar{Q})=P \bar{P} Q (\partial_{\bar\mu} \partial_{\bar\nu}\bar{Q})\partial_m \bar \psi^{\bar\mu} \partial^m \bar \psi^{\bar\nu}$
can be performed, passing from space-time derivatives to field derivatives.

Now substitute $X^{m},\bar X^{m}$ (\ref{st-tq}) and $\tilde X^{m},\bar{\tilde{X}}^{m}$ (\ref{coord1}),
expanding around the space-time coordinates $x^{m}$. Thus, the terms contributing to the K\"ahler potential, those are the ones with 
$\theta^4$ powers in (\ref{l1}):

\begin{eqnarray}
&&\kappa^{\underline{\mu},\underline{\nu},\underline{\rho},\underline{\eta}}|_{\theta^4}=  \label{eqTheta4} \\
&&\frac{1}{4}P \bar{P} Q (\partial_m \partial^m \bar{Q}) -\frac{1}{4}P(\partial_m \partial^m \bar{P}) Q \bar{Q}+ \frac{1}{4}P \bar{P} (\partial_m \partial^m Q) \bar{Q}\nn\\
&&-\frac{1}{4}(\partial_m \partial^m P) \bar{P} Q \bar{Q}+\frac{1}{2}(\partial_m P) (\partial^m \bar{P}) Q \bar{Q}-\frac{1}{2} P \bar{P} (\partial_m Q )(\partial^m \bar{Q}) \nonumber \\
&&+\frac{1}{2} (\partial_m P) \bar{P} (\partial_n Q) \bar{Q} \epsilon^{mn}+\frac{1}{2}P(\partial_m \bar{P}) Q (\partial_n \bar{Q})\epsilon^{mn}\nonumber\\
&&-\frac{1}{2} (\partial_m P) \bar{P} Q (\partial_n \bar{Q}) \epsilon^{mn} -\frac{1}{2} P (\partial_m \bar{P})(\partial_n Q) \bar{Q} \epsilon^{mn} \nonumber \\
&&- \frac{i}{2} \left[\chi^\mu_+ \partial_m(\bar\chi^{\bar\mu}_{\dot{+}} \partial_{\bar\mu} \bar{P}) \sigma^m_{- \dot{-}}+\chi^\mu_- \partial_m(\bar\chi^{\bar\mu}_{\dot{-}} \partial_{\bar\mu} \bar{P}) \sigma^m_{+ \dot{+}} \right]( \partial_\mu P) Q \bar{Q} \nonumber \\
&&+\frac{i}{2} \left[ \partial_m(\chi^\mu_+ \partial_\mu P) \bar\chi^{\bar\mu}_{\dot{+}}\sigma^m_{- \dot{-}} +\partial_m(\chi^\mu_- \partial_\mu P) \bar\chi^{\bar\mu}_{\dot{-}}\sigma^m_{+ \dot{+}} \right] (\partial_{\bar\mu}\bar{P}) Q \bar{Q} \nonumber \\
&&+\frac{i}{2} (\chi^{\mu}_{+} \bar\chi^{\bar{\mu}}_{\dot{+}} \sigma^m_{- \dot{-}}-\chi^{\mu}_{-} \bar\chi^{\bar{\mu}}_{\dot{-}} \sigma^m_{+ \dot{+}}) \left( \partial_\mu P \right) (\partial_{\bar\mu} \bar{P}) Q (\partial_m \bar{Q}) \nonumber \\
&&+\frac{i}{2} \left( \chi^\mu_- \bar\chi^{\bar\mu}_{\dot{-}} \sigma^m_{+ \dot{+}} -\chi^\mu_+ \bar\chi^{\bar\mu}_{\dot{+}} \sigma^m_{- \dot{-}} \right) (\partial_\mu P) (\partial_{\bar\mu} \bar{P}) (\partial_m Q) \bar{Q}\nonumber  \\
&&+\frac{i}{2} \left( (\tilde\chi^{\mu}_{+} \partial_{\mu} Q )\partial_m (\bar{\tilde\chi}^{\bar\mu}_{\dot{+}} \partial_{\bar\mu} \bar{Q})\sigma^m_{- \dot{-}}+ (\tilde\chi^{\mu}_{-} \partial_{\mu} Q ) \partial_m (\bar{\tilde\chi}^{\bar\mu}_{\dot{-}} \partial_{\bar\mu} \bar{Q})\sigma^m_{+\dot{+}} \right) P \bar{P} \nonumber\\
&&-\frac{i}{2}\left( \partial_m (\tilde\chi^{\mu}_{+} \partial_{\mu} Q) (\bar{\tilde\chi}^{\bar\nu}_{\dot{+}} \partial_{\bar\nu} \bar{Q}) \sigma^m_{- \dot{-}}
+ \partial_m (\tilde\chi^{\mu}_{-} \partial_{\mu} Q) (\bar{\tilde\chi}^{\bar\nu}_{\dot{-}} \partial_{\bar\nu} \bar{Q}) \sigma^m_{+ \dot{+}} \right)  P \bar{P} \nonumber\\
&&+ \frac{i}{2} \left( \tilde\chi^{\mu}_{+} \bar{\tilde\chi}^{\bar\mu}_{\dot{+}} \sigma^m_{- \dot{-}}-\tilde\chi^{\mu}_{-} \bar{\tilde\chi}^{\bar\mu}_{\dot{-}} \sigma^m_{+ \dot{+}} \right) (\partial_m P) \bar{P} (\partial_\mu Q) (\partial_{\bar\mu} \bar{Q}) \nonumber \\
&&- \frac{i}{2}\left(\tilde\chi^{\mu}_{+} \bar{\tilde\chi}^{\bar\mu}_{\dot{+}} \sigma^m_{- \dot{-}}-\tilde\chi^{\mu}_{-} \bar{\tilde\chi}^{\bar\mu}_{\dot{-}} \sigma^m_{+ \dot{+}} \right) P (\partial_m \bar P)(\partial_\mu Q) (\partial_{\bar\mu} \bar{Q}) \nonumber \\
 && -\frac{i}{2} \partial_m(\bar{\chi}^{\bar\mu}_{\dot{-}} \partial_{\bar\mu}\bar P) (\bar{\tilde\chi}^{\bar{\nu}}_{\dot{-}}\partial_{\bar\nu} \bar{Q})  PQ \sigma^m_{+ \dot{+}} -\frac{i}{2} (\bar\chi^{\bar\mu}_{\dot{-}}\partial_{\bar\mu} \bar{P}) \partial_m(\bar{\tilde\chi}^{\bar\nu}_{\dot{-}}  \partial_{\bar\nu} \bar{Q})PQ \sigma^m_{+ \dot{+}} \nonumber \\
 &&+\frac{i}{2}(\bar\chi^{\bar\mu}_{\dot{-}} \partial_{\bar\mu} \bar{P}) (\bar{\tilde\chi}^{\bar\nu}_{\dot{-}}\partial_{\bar\nu} \bar{Q}) (\partial_m P)Q \sigma^m_{+ \dot{+}} +\frac{i}{2} (\bar{\chi}^{\bar\mu}_{\dot{-}} \partial_{\bar\mu} \bar{P}) (\bar{\tilde\chi}^{\bar{\nu}}_{\dot{-}}\partial_{\bar\nu} \bar{Q}) P(\partial_m Q) \sigma^m_{+ \dot{+}} \nonumber \\
 &&-\frac{i}{2} \partial_m(\chi^\mu_- \partial_\mu P) (\tilde\chi^{\nu}_{-} \partial_{\nu} Q)\bar{P}\bar{Q} \sigma^m_{+ \dot{+}}-\frac{i}{2}(\chi^{\mu}_{-}\partial_\mu P)\partial_m (\tilde\chi^\nu_{-} \partial_\nu Q) \bar{P} \bar{Q } \sigma^m_{+ \dot{+}} \nonumber\\
 && + \frac{i}{2} (\chi^\mu_{-} \partial_\mu P) (\tilde\chi^\nu_{-} \partial_\nu Q) (\partial_m \bar{P}) \bar{Q} \sigma^m_{+ \dot{+}}+\frac{i}{2}(\chi^\mu_{-} \partial_{\mu} P)(\tilde\chi^{\nu}_{-}\partial_{\nu} Q)  \bar{P} \partial_m\bar{Q}\sigma^m_{+ \dot{+}}
\nonumber 
\end{eqnarray}
\begin{eqnarray}
&&+
\frac{i}{2} \partial_m (\chi^{\mu}_{+} \partial_\mu P) (\bar{\tilde\chi}^{\bar\mu}_{\dot{+}}  \partial_{\bar{\mu}} \bar{Q} ) Q \bar{P}\sigma^m_{- \dot{-}}+\frac{i}{2} (\chi^{\mu}_{+}\partial_\mu P) \partial_m(\bar{\tilde\chi}^{\bar\mu}_{\dot{+}} \partial_{\bar\mu} \bar{Q}) Q  \bar{P} \sigma^m_{- \dot{-}}  \nonumber \\
&&- \frac{i}{2} (\chi^{\mu}_{+}  \partial_\mu P)(\bar{\tilde\chi}^{\bar\mu}_{\dot{+}} \partial_{\bar\mu} \bar{Q}) Q (\partial_m \bar{P}) \sigma^m_{- \dot{-}}- \frac{i}{2} (\chi^\mu_{+} \partial_\mu P)  (\bar{\tilde\chi}^{\bar\mu}_{\dot{+}} \partial_{\bar\mu} \bar{Q})\bar{P} (\partial_m Q)\sigma^m_{- \dot{-}} \nonumber \\
&&  +\frac{i}{2}  \partial_{m}(\bar\chi^{\bar\mu}_{\dot{+}}\partial_{\bar\mu} \bar{P} ) ( \tilde\chi^{\mu}_{+} \partial_\mu Q) \bar{Q} P \sigma^m_{-\dot{-}}  +\frac{i}{2}  (\bar\chi^{\bar\mu}_{\dot{+}}\partial_{\bar\mu} \bar{P}) \partial_{m} ( \tilde\chi^{\mu}_{+} \partial_\mu Q) \bar{Q} P \sigma^m_{-\dot{-}}  \nonumber \\
&&   - \frac{i}{2}  (\bar{\chi}^{\bar\mu}_{\dot{+}}\partial_{\bar{\mu}} \bar{P}) (\tilde\chi^{\mu}_{+}  \partial_\mu Q)  (\partial_m P) \bar{Q} \sigma^m_{- \dot{-}}-\frac{i}{2}  (\bar\chi^{\bar\mu}_{\dot{+}}\partial_{\bar\mu} \bar{P}) (\tilde\chi^{\nu}_{+} \partial_\nu Q) P  (\partial_m \bar{Q}) \sigma^m_{-\dot{-}} \nonumber \\
&&+ \chi^\mu_{-} \bar{\chi}^{\bar\mu}_{\dot{+}} (\partial_\mu P) (\partial_{\bar\mu} \bar{P}) g \bar{Q}  + \chi^{\mu}_{+} \bar\chi^{\bar{\mu}}_{\dot{-}}(\partial_\mu P)(\partial_{\bar\mu} \bar{P}) \bar{g} Q \nonumber \\
&& - \tilde\chi^{\mu}_{-} \bar{\tilde\chi}^{\bar\mu}_{\dot{+}} f \bar{P} (\partial_{\mu} Q) (\partial_{\bar\mu} \bar{Q})- \tilde\chi^\mu_{+} \bar{\tilde\chi}^{\bar\mu}_{\dot{-}} (\partial_\mu Q) \bar{f} P (\partial_{\bar\mu} \bar{Q}) \nonumber\\
&&-\chi^\mu_{+} \bar{\tilde\chi}^{\bar\mu}_{\dot{-}} (\partial_\mu P)\bar{f} Q (\partial_{\bar\mu} \bar{Q})+ \bar\chi^{\bar\mu}_{\dot{+}} \tilde\chi^{\mu}_{-} f (\partial_{\bar\mu} \bar{P}) (\partial_\mu Q) \bar{Q}\nonumber \\
&& + \bar\chi^{\bar\mu}_{\dot{-}} \bar{\tilde\chi}^{\bar\nu}_{\dot{+}} f (\partial_{\bar\mu} \bar{P}) Q (\partial_{\bar\nu} \bar{Q})- \chi^{\mu}_{-} \tilde\chi^{\nu}_{+} \bar f (\partial_{\mu} P) \bar Q (\partial_{\nu} Q)  \nonumber \\
&&+  \bar\chi^{\bar\mu}_{\dot{+}} \bar{\tilde\chi}^{\bar\nu}_{\dot{-}}P (\partial_{\bar\mu} \bar{P}) g (\partial_{\bar\nu} \bar{Q})-  
\chi^{\mu}_{+} \tilde\chi^{\nu}_{-}\bar P (\partial_{\mu} P) \bar g (\partial_{\nu} Q)  \nonumber \\
&& +\chi^{\mu}_{-} \bar{\tilde\chi}^{\bar\nu}_{\dot+} (\partial_{\mu} P) (\partial_{\bar\nu} \bar{Q}) \bar{P} g  - \bar\chi^{\bar\mu}_{\dot-} \tilde\chi^{\nu}_{+} (\partial_{\bar\mu} \bar{P}) (\partial_{\nu} Q) P \bar{g}  \\
&&+(\chi^{\mu}_{-} \bar\chi^{\bar\mu}_{\dot{-}} \tilde\chi^{\nu}_{+} \bar{\tilde\chi}^{\bar\nu}_{\dot{+}}-\chi^{\mu}_{+} \bar\chi^{\bar\mu}_{\dot{+}} \tilde\chi^{\nu}_{-} \bar{\tilde\chi}^{\bar\nu}_{\dot{-}}) (\partial_\mu P)(\partial_{\bar\mu} \bar{P})(\partial_{\nu} Q )(\partial_{\bar\nu} \bar{Q})\nonumber \\
&&-P \bar{P} g \bar{g} +Q \bar{Q} f \bar{f}  \nonumber 
\end{eqnarray}
Let us integrate by parts and arrange one of the contributions from $(\ref{eqTheta4})$. Substituting explicitly $\bar{f}$ and $\bar{g}$, 
and with the aid of the definition $(\ref{metric})$ we find

\begin{eqnarray}
\kappa^{\underline{\mu},\underline{\nu},\underline{\rho},\underline{\eta}}|_{\theta^4}&&\supset
 \left[ (\partial_m P) (\partial^m \bar{P}) Q \bar{Q} - P \bar{P} (\partial_m Q) (\partial^m \bar{Q})\right] \label{L2} \\
&&+ \left[ P(\partial_m \bar{P})(\partial_n Q) \bar{Q} + (\partial_m P) \bar{P} Q (\partial_n \bar{Q}) \right] \epsilon^{mn} \nonumber \\
&&+ \left[F^{\mu} \partial_{\mu} P - \frac{1}{2} \chi^{\mu} \chi^{\nu} \partial_{\mu} \partial_{\nu} P \right] \left[\bar{F}^{\bar\mu} \partial_{\bar\mu} \bar{P} - \frac{1}{2} \bar{\chi}^{\bar\mu} \bar{\chi}^{\bar\nu} \partial_{\bar\mu} \partial_{\bar\nu} \bar{P} \right] Q \bar{Q} \nonumber \\
&&-  P \bar{P} \left[G^{\mu} \partial_{\mu} Q - \frac{1}{2} \tilde\chi^{\mu} \tilde\chi^{\nu} \partial_{\mu} \partial_{\nu} Q \right]\left[\bar{G}^{\bar\mu} \partial_{\bar\mu} \bar{Q} - \frac{1}{2} \bar{\tilde\chi}^{\bar\mu} \bar{\tilde\chi}^{\bar\nu} \partial_{\bar\mu} \partial_{\bar\nu} \bar{Q} \right]. \nn
\end{eqnarray}
From this expression we will cast the purely bosonic contributions to the
K\"ahler potential. Now we cast the whole sum in $(\ref{kaehlerTot})$, using the linearity of the
calculation to collect all the terms contributing to a generic Lagrangian $(\ref{LTot})$. Let us
consider $k=\sum_{i,j,m,n}\sum_{\underline{\mu},\underline{\nu},\underline{\rho},\underline{\eta}} c_{\underline{\mu},\underline{\nu},\underline{\rho},\underline{\eta}}  P \bar P Q \bar Q$ as the sum of all terms in the series of the K\"ahler potential $(\ref{kaehlerTot})$ evaluated in
the scalars and write the definitions
\begin{eqnarray}
k_{\mu_1 \bar\nu_1}&=&\frac{\partial^2}{\partial \phi^{\mu_1} \partial \bar \phi^{\nu_1}} k, \, \, k_{\mu_2 \bar\nu_2}=\frac{\partial^2}{\partial \psi^{\mu_2} \partial \bar \psi^{\nu_2}} k, \label{kaehlers}\\
k_{\mu_1 \bar\nu_2}&=&\frac{\partial^2}{\partial \phi^{\mu_1} \partial \bar \psi^{\nu_1}} k, \, \, k_{\mu_2 \bar\nu_1}=\frac{\partial^2}{\partial \psi^{\mu_2} \partial \bar \phi^{\nu_1}} k, \nn \\ 
k_{\mu \bar\nu}&=& \{k_{\mu_1 \bar\nu_1},k_{\mu_1 \bar\nu_2},k_{\mu_2 \bar\nu_1},k_{\mu_2 \bar\nu_2}\}, \nn\\
\label{chris2}
\Gamma^\sigma_{\mu \rho}&=&g^{\sigma \bar\nu}\frac{\partial}{\partial A^\rho} g_{\mu \bar\nu} , \, \Gamma^{\bar \sigma}_{\bar\nu \bar\rho}= g^{\mu \bar \sigma}\frac{\partial}{\partial \bar A^\rho} g_{\mu \bar \nu}, \, \, A^{\rho}=(\phi^{\rho_1}, \psi^{\rho_2}). \nn
\end{eqnarray}
We have denoted the K\"ahler metric by $k_{\mu \bar\nu}$. We consider the following convention, the greek indices with a $1$ subindex: $\mu_1, \nu_1,\ldots$ are related to the chiral scalars component $\phi$. Similarly those related to the twisted chiral scalar component $\psi$ are denoted with greek indices
and a subindex $2$: $\mu_2, \nu_2,\ldots$. We will denote the greek indices
without subindices as $\rho=(\rho_1,\rho_2)$ running over both scalar components of
chiral and twisted chiral superfields.

The sum of all the terms like (\ref{L2}) can be arranged by using equations (\ref{kaehlers}) i.e. giving
for the total K\"ahler potential $K= \sum_{i,j,m,n}\sum_{\underline{\mu},\underline{\nu},\underline{\rho},\underline{\eta}} c_{\underline{\mu},\underline{\nu},\underline{\rho},\underline{\eta}} \kappa^{\underline{\mu},\underline{\nu},\underline{\rho},\underline{\eta}}$,
giving

\begin{eqnarray}
\label{bos}
K|_{\theta^4}&\supset& \left[ (\partial_m \phi^{\mu_1}) (\partial^m \bar{\phi}^{\bar{\mu}_1}) k_{\mu_1 \bar\mu_1} - (\partial_m \psi^{\mu_2}) (\partial^m \bar{\psi}^{\bar\mu_2}) k_{\mu_2 \bar{\mu}_2}\right] \\
&&+ \left[ (\partial_m \bar\phi^{\bar\mu_1}) (\partial_n \psi^{\mu_2}) k_{\mu_2 \bar\mu_1} + (\partial_m \phi^{\mu_1}) (\partial_n \bar\psi^{\bar\mu_2}) k_{\mu_1 \bar\mu_2} \right] \epsilon^{mn} \nonumber  \\
&&+ \left[F^{\mu_1} \bar{F}^{\bar\mu_1} k_{\mu_1 \bar\mu_1} - \frac{1}{2} \bar\chi^{\bar\mu_1} \bar\chi^{\bar\nu_1} F^{\mu_1} k_{\mu_1 \bar\rho} \Gamma^{\bar\rho}_{\bar\mu_1 \bar\nu_1} \right. \nonumber \\ 
&&- \left. \frac{1}{2} \chi^{\mu_1} \chi^{\nu_1} \bar{F}^{\bar\mu_1} k_{\rho \bar\mu_1} \Gamma^{\rho}_{\mu_1 \nu_1} + \frac{1}{4} \chi^{\mu_1} \chi^{\nu_1} \bar{\chi}^{\bar\mu_1} \bar{\chi}^{\bar\nu_1} (\partial_{\nu_1 \bar\nu_1} k_{\mu_1 \bar{\mu}_1}) \right] \nonumber 
\\
&&- \left[G^{\mu_2} \bar{G}^{\bar\mu_2} k_{\mu_2 \bar\mu_2} - \frac{1}{2} \bar{\tilde\chi}^{\bar\mu_2} \bar{\tilde\chi}^{\bar\nu_2} G^{\mu_2} k_{\mu_2 \bar\rho} \Gamma^{\bar\rho}_{\bar\mu_2 \bar\nu_2} \right. \nonumber \\ 
&&- \left. \frac{1}{2} \tilde\chi^{\mu_2} \tilde\chi^{\nu_2} \bar{G}^{\bar\mu_2} k_{\rho \bar\mu_2} \Gamma^{\rho}_{\mu_2 \nu_2} + \frac{1}{4} \tilde\chi^{\mu_2} \tilde\chi^{\nu_2} \bar{\tilde\chi}^{\bar\mu_2} \bar{\tilde\chi}^{\bar\nu_2} (\partial_{\nu_2 \bar\nu_2} k_{\mu_2 \bar{\mu}_2}) \right].\nonumber
\end{eqnarray}
Thus the purely bosonic part (including the auxiliary fields) is given by
\begin{eqnarray}
K|_{\theta^4}&\supset& \left[ (\partial_m \phi^{\mu_1}) (\partial^m \bar{\phi}^{\bar{\mu}_1}) k_{\mu_1 \bar\mu_1} - (\partial_m \psi^{\mu_2}) (\partial^m \bar{\psi}^{\bar\mu_2}) k_{\mu_2 \bar{\mu}_2}\right] \\
&&+ \left[ (\partial_m \bar\phi^{\bar\mu_1}) (\partial_n \psi^{\mu_2}) k_{\mu_2 \bar\mu_1} + (\partial_m \phi^{\mu_1}) (\partial_n \bar\psi^{\bar\mu_2}) k_{\mu_1 \bar\mu_2} \right] \epsilon^{mn} \nonumber \\
&&+ F^{\mu_1} \bar{F}^{\bar\mu_1} k_{\mu_1 \bar\mu_1}- G^{\mu_2} \bar{G}^{\bar\mu_2} k_{\mu_2 \bar\mu_2}. \nn
\end{eqnarray}
The scalar contributions to the Lagrangian in the first line reproduce the ones given in \cite{rocek84}. The contributions to the Lagrangian which include fermions from (\ref{eqTheta4}) are listed in formula (\ref{bos}) from Appendix \ref{two}.

\section{Final Remarks}
\label{seven}

In these notes we develop a method to describe
the superspace coordinates in Lagrangians where twisted superfields are present,
and we discuss some of the applications. The chosen coordinates allow to obtain
all kinetic terms and interactions of arbitrary Lagrangians depending on twisted
chiral superfields. We discuss as examples the Abelian T-dual model to a 
single chiral superfield $U(1)$ GLSM, and the non-Abelian T-dual model to the
$\mathbb{CP}^1$ $U(1)$ GLSM. We also apply the technique to study models with both chiral and
twisted chiral multiplets. The method developed could be useful in future investigations of 
T-dualities of Gauged Linear Sigma Models and 
Mirror Symmetry for target manifolds. 



The approach employed consists in treating the twisted chiral superfields Lagrangians
analogously to the way chiral superfields Lagrangians are treated. This is
intuitive considering the permutation symmetry that exists between both
representations. We define superspace coordinates in a convenient scheme, this is
the redefined space-time coordinates are annihilated by covariant derivatives $\bar D_+$
and $D_-$. This allows to determine the most general twisted superfields expansions.
Those modified coordinates allow to perform the expansions of the K\"ahler potential and
twisted superpotential in a sistematic way, collecting all
the contributions to the Lagrangian.

We have also considered generic Lagrangian with both twisted chiral
and chiral representations. In those calculations we have employed a mixture of both
space-time coordinates redefinitions (twisted or not), convenient to
perform the expansions of (twisted-) chiral superfields.
Such Lagrangians are more exotic. However, when one considers generic 
gauged linear sigma models in 2D with gauged global symmetries, giving
rise to master Lagrangians to explore generic T-dualities \cite{CaboBizet:2017fzc},
those master Lagrangians posses both: twisted chiral
and chiral superfields. Thus even this approach could find
eventually an application in the study of two dimensional sigma models.


We intend to apply our results to compute the sphere partition functions of non-Abelian T-duals from
gauged linear sigma models  \cite{CaboBizet:2017fzc}. The partition function of the original
gauged linear sigma models with chiral superfields has already been studied \cite{gomis1,benini1}. The matching
of the partition function of the dual models
could serve to establish the duality in more solid grounds. 
In those computations the fermionic components of the action are relevant
for the use of supersymmetric localization techniques \cite{gomis1,benini1,doroud1,doroud2}. The Lagrangian of
the dual models depend on twisted chiral superfields, from which we have determined here the fermionic contributions. There could be connections between Mirror
Symmetry in non complete intersection Calabi-Yau varieties \cite{Jockers:2012zr,Gu:2018fpm,Chen:2018wep,Honma:2018fgw} and 
non-Abelian T-dualities, where twisted superfields arise. The methods studied here can be used to
explore these topics.

\section{Acknowledgments}
We thank Leopoldo Pando-Zayas for helpful discussions and useful comments on the manuscript. 
We thank Nima Doroud, Hans Jockers, Yulier Jim\'enez Santana,  Albrecht Klemm, Aldo Mart\'{\i}nez-Merino, Imtak Jeon and Kumar Narain for useful discussions. We thank the Casa Matem\'atica de Oaxaca, BIRS, and the organizers for the workshop ``Geometrical tools for string cosmology"
for a beautiful scientific environment. NCB thanks the support of the project CIIC 181/2019 of the University of Guanajuato and CONACYT Project A1-S-37752, she also thanks ICTP HECAP Section. 
RSS thanks the support of PRODEP NPTC Project UDG-PTC-1368.

\appendix

\section{Notation and Conventions}
\label{one}
In this Appendix we set our conventions to denote spinors, superspace coordinates and spinor products which are the same as the one presented in \cite{wessbagger}.

The Grassman superspace variables anti-commute which each other i.e. let $\theta_1$ and $\theta_2$ Grassman variables they satisfy $\theta_1 \theta_2$ = - $\theta_2 \theta_1$. The fermionic fields and their adjoint conjugates are given by
\begin{equation}
\psi = \left(\begin{array}{c} \psi_\alpha \\ \bar\psi^{\dot\alpha} \end{array}\right) \quad \psi^{\dag} = \left(\begin{array}{cc} \bar\psi_{\dot\alpha} & \psi^{\alpha} \end{array}\right).
\end{equation}
The component spinors considered are Weyl.  The ones with dotted (undotted) indices transform on the $(0,\frac{1}{2})$ $((\frac{1}{2},0))$ representation of the Lorentz group. The supersymmetric covariant derivatives are given by
\begin{eqnarray}
D_{\alpha} = \frac{\partial}{\partial \theta^{\alpha}} + i \bar\theta^{\dot\alpha} \sigma_{\alpha \dot\alpha}^m \partial_m, \\
\bar{D}_{\dot\alpha} = -\frac{\partial}{\partial \bar\theta^{\dot\alpha}} -i \theta^{\alpha} \sigma_{\alpha \dot\alpha}^m \partial_m.
\end{eqnarray}
We use Levi-Civitta symbols as
\begin{equation}
\epsilon^{\alpha \beta}= \epsilon^{\dot\alpha \dot\beta} = \left(\begin{array}{cc} 0 & 1 \\ -1 & 0 \end{array}\right), \quad \quad \epsilon_{\alpha \beta}= \epsilon_{\dot\alpha \dot\beta} = \left(\begin{array}{cc} 0 & -1 \\ 1 & 0 \end{array}\right). 
\end{equation}
They satisfy the identity $\epsilon_{\alpha \beta} \epsilon^{\beta \gamma} = \delta^\gamma_\alpha$. Moreover we have
the following properties for fermion products
\begin{eqnarray}
\psi \chi &=& \psi^\alpha \chi_\alpha = - \psi_\alpha \chi^\alpha , \quad \quad \bar\psi \bar\chi = \bar\psi_{\dot\alpha} \bar\chi^{\dot\alpha} = - \bar\psi^{\dot\alpha} \chi_{\dot\alpha},\\
\psi^\alpha&=& \epsilon^{\alpha \beta} \psi_\beta, \quad \quad \psi_{\alpha} =\epsilon_{\alpha \beta} \psi^\beta.
\end{eqnarray}
Now we define the $\sigma$'s matrices as follows

\begin{eqnarray}
\sigma_0 = \left(\begin{array}{cc} -1 & 0 \\ 0 & -1 \end{array}\right) ,\quad \sigma_1 = \left(\begin{array}{cc} 0 & 1 \\ 1 & 0 \end{array}\right), \nonumber \\
\sigma_2 = \left(\begin{array}{cc} 0 & -i \\ i & 0 \end{array}\right), \quad \sigma_3 = \left(\begin{array}{cc} 1 & 0 \\ 0 & -1 \end{array}\right),
\end{eqnarray}
1, 2 and 3 are the Pauli matrices. Along this paper we consider the following conventions for products of Grassman superspace coordinates

\begin{eqnarray}
\theta^2 &=& \theta^{\alpha} \theta_{\alpha} = \theta^+ \theta_+ + \theta^- \theta_- = - 2 \theta^+ \theta^-, \\ \bar\theta^2 &=& \bar\theta_{\dot\alpha} \bar\theta^{\dot\alpha} = \bar\theta_{\dot+} \bar\theta^{\dot +} + \bar\theta_{\dot -} \bar\theta^{\dot -} = 2 \bar\theta^{\dot +} \bar\theta^{\dot -}, \\
\theta^{\alpha} \theta^{\beta} &=& -\frac{1}{2} \epsilon^{\alpha \beta} \theta^2, \quad \quad \bar\theta^{\dot\alpha} \bar\theta^{\dot\beta} = \frac{1}{2} \epsilon^{\dot\alpha \dot\beta} \bar\theta^2.
\end{eqnarray}
Let us add the space-time redefinition coordinates, which serve to express the expansions 
of chiral superfields:
\begin{eqnarray}
X^m& =& x^m + i \theta^\alpha \sigma_{\alpha \dot{\beta}}^m \bar\theta^{\dot{\beta}}. \label{coord1} 
\end{eqnarray}
They are annihilated by $\bar D_{\pm}$. Their conjugated are annihilated by $D_{\pm}$.

Now let us introduce a set of identities useful for the computation of the K\"ahler potential of twisted and
chiral superfield.
\begin{eqnarray}
\theta^{\alpha}\theta^{\beta}&=&-\frac{1}{2}\epsilon^{\alpha\beta}\theta^2,  \, \,
{\tilde\theta}^{\alpha}\tilde\theta^{\beta}=-\frac{1}{2}\epsilon^{\alpha\beta}{\tilde\theta}^2, \\
\bar\theta^{\alpha}\bar\theta^{\beta}&=&\frac{1}{2}\epsilon^{\alpha \beta}\bar\theta^2,  \,   \,
\bar{\tilde\theta}^{\alpha}\bar{\tilde\theta}^{\beta}=\frac{1}{2}\epsilon^{\dot\alpha\dot\beta}\bar{\tilde\theta}^2,\nonumber
\end{eqnarray}

\begin{eqnarray}
\theta^2 (\theta \sigma \bar \theta)&=& \tilde \theta^2 (\theta  \sigma \bar \theta)= \theta^2 (\tilde\theta \sigma \bar {\tilde\theta})=  \tilde \theta^2 (\tilde\theta \sigma \bar {\tilde\theta})=0, \\
    \tilde \theta^2 \theta^2&=& \tilde \theta^2 \bar\theta^2=\bar{\tilde\theta}^2 {\bar \theta}^2= \bar{\tilde\theta}^2 { \theta}^2=0. \nonumber
\end{eqnarray}

\begin{eqnarray}
(\tilde\theta \sigma^m \bar{\tilde\theta})(\tilde\theta \sigma^n \bar{\tilde\theta})&=&-2 \bar{\theta}^+\bar{\theta}^-\theta^-\theta^+\eta^{mn},\\\
(\theta \sigma^m \bar{\theta})(\theta \sigma^n \bar{\theta})&=&2 \bar{\theta}^+\bar{\theta}^-\theta^-\theta^+\eta^{mn},\nonumber \\
(\theta \sigma^m \bar{\theta})(\tilde\theta \sigma^n \bar{\tilde\theta})&=&-2 \bar{\theta}^+\bar{\theta}^-\theta^-\theta^+\epsilon^{mn},\nonumber\\
\theta^{\alpha}\bar{\theta}^{\dot \tau}(\theta^{\beta} \sigma^m_{\beta \dot \gamma} \bar{\theta}^{\dot \gamma})&=& \bar{\theta}^+\bar{\theta}^-\theta^-\theta^+ (\bar \sigma^m)^{\dot\tau \alpha}, \nn \\
\tilde\theta^{\alpha}\bar{\tilde\theta}^{\dot \tau}({\tilde\theta}^{\beta} \sigma^m_{\beta \dot \gamma} \bar{\tilde\theta}^{\dot \gamma})&=&- \bar{\theta}^+\bar{\theta}^-\theta^-\theta^+ (\bar \sigma^m)^{\dot\tau \alpha}.\nonumber
\end{eqnarray}
Also we employ the following redefinition:

\begin{eqnarray}
(\bar\sigma^{m})^{\dot \tau \alpha}=\epsilon^{\dot \tau \dot\gamma}\epsilon^{\alpha \beta} \sigma^m_{\beta \dot \gamma},\bar\sigma^0=\sigma_0, 
\bar\sigma^3=-\sigma_3. 
\end{eqnarray}
Employing these definitions we can rewrite the lines $6-7$ and $10-11$ from equation $(\ref{eqTheta4})$ in a more compact form:

\begin{eqnarray}
&&{\frac{i}{2} (\bar\sigma^m)^{\dot\tau \alpha}\left[ \partial_m(\chi^\mu_{\alpha} \partial_\mu P) \bar\chi^{\bar\mu}_{\dot{\tau}} \partial_{\bar\mu}\bar{P} -(\chi^\mu_{\alpha} \partial_\mu P) \partial_m(\bar\chi^{\bar\mu}_{\dot{\tau}}\partial_{\bar\mu}\bar P) \right] Q \bar{Q}} = \\
 &&- \frac{i}{2} \left[\chi^\mu_+ \partial_m(\bar\chi^{\bar\mu}_{\dot{+}} \partial_{\bar\mu} \bar{P}) \sigma^m_{- \dot{-}}+\chi^\mu_- \partial_m(\bar\chi^{\bar\mu}_{\dot{-}} \partial_{\bar\mu} \bar{P}) \sigma^m_{+ \dot{+}} \right]( \partial_\mu P) Q \bar{Q} \nonumber \\
&&+\frac{i}{2} \left[ \partial_m(\chi^\mu_+ \partial_\mu P) \bar\chi^{\bar\mu}_{\dot{+}}\sigma^m_{- \dot{-}} +\partial_m(\chi^\mu_- \partial_\mu P) \bar\chi^{\bar\mu}_{\dot{-}}\sigma^m_{+ \dot{+}} \right] (\partial_{\bar\mu}\bar{P}) Q \bar{Q} ,\nonumber \\
&&{-\frac{i}{2} (\bar\sigma^m)^{\dot\tau \alpha}\left[ \partial_m(\tilde\chi^\mu_{\alpha} \partial_\mu Q) \bar{\tilde\chi}^{\bar\mu}_{\dot{\tau}} \partial_{\bar\mu}\bar{Q} -({\tilde\chi}^\mu_{\alpha} \partial_\mu Q) \partial_m(\bar{\tilde\chi}^{\bar\mu}_{\dot{\tau}}\partial_{\bar\mu}\bar Q) \right] P \bar{P}}=\\
&&+\frac{i}{2} \left( (\tilde\chi^{\mu}_{+} \partial_{\mu} Q )\partial_m (\bar{\tilde\chi}^{\bar\mu}_{\dot{+}} \partial_{\bar\mu} \bar{Q})\sigma^m_{- \dot{-}}+ (\tilde\chi^{\mu}_{-} \partial_{\mu} Q ) \partial_m (\bar{\tilde\chi}^{\bar\mu}_{\dot{-}} \partial_{\bar\mu} \bar{Q})\sigma^m_{+\dot{+}} \right) P \bar{P} \nonumber\\
&&-\frac{i}{2}\left( \partial_m (\tilde\chi^{\mu}_{+} \partial_{\mu} Q) (\bar{\tilde\chi}^{\bar\nu}_{\dot{+}} \partial_{\bar\nu} \bar{Q}) \sigma^m_{- \dot{-}}
+ \partial_m (\tilde\chi^{\mu}_{-} \partial_{\mu} Q) (\bar{\tilde\chi}^{\bar\nu}_{\dot{-}} \partial_{\bar\nu} \bar{Q}) \sigma^m_{+ \dot{+}} \right)  P \bar{P}. \nonumber
 \end{eqnarray}

\section{Fermionic Contribution Chiral and Twisted Chiral Action}
\label{two}

In this Appendix we write explicitly all the fermionic terms of the K\"ahler potential
for a generic Lagrangian of twisted and chiral superfields coupled:

\begin{eqnarray}
&K|_{\theta^4}& \supset -\frac{i}{2} \chi^{\mu_1}_{+}  \left[ (\partial_m \bar\chi^{\bar\mu_1}_{\dot{+}}) k_{\mu_1 \bar\mu_1} + \bar\chi^{\bar\mu_1}_{\dot{+}} (\partial_m \bar\phi^{\bar\nu_1}) k_{\mu_1 \bar\rho} \Gamma^{\bar\rho}_{\bar\mu_1 \bar\nu_1} \right] \sigma^m_{- \dot{-}}\nonumber \\
 & &-\frac{i}{2} \chi^{\mu_1}_{-} \left[ (\partial_m \bar\chi^{\bar\mu_1}_{\dot{-}} ) k_{\mu_1 \bar\mu_1} + \bar\chi^{\bar\mu_1}_{\dot{-}} (\partial_m \bar\phi^{\bar\nu_1})  k_{\mu_1 \bar\rho} \Gamma^{\bar\rho}_{\bar\mu_1 \bar\nu_1}\right] \sigma^m_{+ \dot{+}} \nonumber \\ 
& & + \frac{i}{2} \left[  (\partial_m \chi^{\mu_1}_{+}) k_{\mu_1 \bar\mu_1} +\chi^{\mu_1}_{+}  (\partial_m \phi^{\nu_1}) k_{\rho \bar\mu_1 } \Gamma^{\rho}_{\mu_1 \nu_1} \right] \bar\chi^{\bar\mu_1}_{\dot{+}} \sigma^m_{- \dot{-}}\nonumber \\
 & &+ \frac{i}{2} \left[  (\partial_m \chi^{\mu_1}_{-}) k_{\mu_1 \bar\mu_1} +\chi^{\mu_1}_{-}  (\partial_m \phi^{\nu_1}) k_{\rho \bar\mu_1 } \Gamma^{\rho}_{\mu_1 \nu_1} \right] \bar\chi^{\bar\mu_1}_{\dot{-}} \sigma^m_{+ \dot{+}} \nonumber \\ 
 & & +  \frac{i}{2} \left[  \chi^{\mu_1}_{+} \bar\chi^{\bar\mu_1}_{\dot{+}}  \sigma^m_{- \dot{-}} -  \chi^{\mu_1}_{-} \bar\chi^{\bar\mu_1}_{\dot{-}}  \sigma^m_{+ \dot{+}} \right] \left[ (\partial_m \bar\psi^{\bar\mu_2}) k_{\mu_1 \bar\rho} \Gamma^{\bar\rho}_{\bar\mu_1 \bar\mu_2} - (\partial_m \psi^{\mu_2}) k_{\rho \bar\mu_1} \Gamma^{\rho}_{\mu_1 \mu_2} \right]  \nonumber \\ 
& & \frac{i}{2} \tilde\chi^{\mu_2}_{+}  \left[ (\partial_m \bar{\tilde\chi}^{\bar\mu_2}_{\dot{+}}) k_{\mu_2 \bar\mu_2} + \bar{\tilde\chi}^{\bar\mu_2}_{\dot{+}} (\partial_m \bar\psi^{\bar\nu_2}) k_{\mu_2 \bar\rho} \Gamma^{\bar\rho}_{\bar\mu_2 \bar\nu_2} \right] \sigma^m_{- \dot{-}} \nonumber \\
 & &\frac{i}{2} \tilde\chi^{\mu_2}_{-} \left[ (\partial_m \bar{\tilde\chi}^{\bar\mu_2}_{\dot{-}} ) k_{\mu_2 \bar\mu_2} + \bar{\tilde\chi}^{\bar\mu_2}_{\dot{-}} (\partial_m \bar\psi^{\bar\nu_2})  k_{\mu_2 \bar\rho} \Gamma^{\bar\rho}_{\bar\mu_2 \bar\nu_2} \right] \sigma^m_{+ \dot{+}} \nonumber \\ 
 & & - \frac{i}{2} \left[  (\partial_m \tilde\chi^{\mu_2}_{+}) k_{\mu_2 \bar\mu_2} + \tilde\chi^{\mu_2}_{+}  (\partial_m \psi^{\nu_2}) k_{\rho \bar\mu_2 } \Gamma^{\rho}_{\mu_2 \nu_2} \right] \bar{\tilde\chi}^{\bar\mu_2}_{\dot{+}} \sigma^m_{- \dot{-}}\nonumber \\
 & &- \frac{i}{2} \left[  (\partial_m \tilde\chi^{\mu_2}_{-}) k_{\mu_2 \bar\mu_2} +\tilde\chi^{\mu_2}_{-}  (\partial_m \psi^{\nu_2}) k_{\rho \bar\mu_2} \Gamma^{\rho}_{\mu_2 \nu_2} \right] \bar{\tilde\chi}^{\bar\mu_2}_{\dot{-}} \sigma^m_{+ \dot{+}} \nonumber \\ 
  & & +  \frac{i}{2} \left[  \tilde\chi^{\mu_2}_{+} \bar{\tilde\chi}^{\bar\mu_2}_{\dot{+}}  \sigma^m_{- \dot{-}} -  \tilde\chi^{\mu_2}_{-} \bar{\tilde\chi}^{\bar\mu_2}_{\dot{-}}  \sigma^m_{+ \dot{+}} \right] \left[ (\partial_m \phi^{\mu_1}) k_{\rho \bar\mu_2} \Gamma^{\rho}_{\mu_2 \mu_1} - (\partial_m \bar\phi^{\bar\mu_1}) k_{\mu_2 \bar\rho} \Gamma^{\bar\rho}_{\bar\mu_2 \bar\mu_1} \right]  \nonumber \\ 
 & & + i \bar\chi^{\bar\mu_1}_{\dot{-}} \bar{\tilde\chi}^{\bar\mu_2}_{\dot{-}} \left[ (\partial_m \phi^{\mu_1}) k_{\mu_1 \bar\rho} + (\partial_m \psi^{\mu_2}) k_{\mu_2 \bar\rho} \right] \Gamma^{\bar\rho}_{\bar\mu_1 \bar\mu_2} \sigma^m_{+ \dot{+}}   \nonumber \\ 
& & + i \chi^{\mu_1}_{-} \tilde\chi^{\mu_2}_{-} \left[ (\partial_m \bar\phi^{\bar\mu_1}) k_{\rho \bar\mu_1} + (\partial_m \bar\psi^{\bar\mu_2}) k_{\rho \bar\mu_2} \right] \Gamma^{\rho}_{\mu_2 \mu_1} \sigma^m_{+ \dot{+}}   \nonumber \\ 
& & - i \chi^{\mu_1}_{+} \bar{\tilde\chi}^{\bar\mu_2}_{\dot{+}} \left[ (\partial_m \bar\phi^{\bar\mu_1}) k_{\mu_1 \bar\rho} \Gamma^{\bar\rho}_{\bar\mu_1 \bar\mu_2}+ (\partial_m \psi^{\mu_2}) k_{\rho \bar\mu_2}  \Gamma^{\rho}_{\mu_2 \mu_1}\right] \sigma^m_{- \dot{-}}   \nonumber \\ 
& & - i \bar\chi^{\bar\mu_1}_{\dot{+}} \tilde\chi^{\mu_2}_{\dot{+}} \left[ (\partial_m \phi^{\mu_1}) k_{\rho \bar\mu_1} \Gamma^{\rho}_{\mu_1 \mu_2}+ (\partial_m \bar\psi^{\bar\mu_2}) k_{\mu_2 \bar\rho}  \Gamma^{\bar\rho}_{\bar\mu_2 \bar\mu_1}\right] \sigma^m_{- \dot{-}}   \nonumber \\ 
& &+ F^{\mu_1} \left[ \bar{\tilde\chi}^{\bar\mu_2}_{\dot{+}} \tilde\chi^{\mu_2}_{-} k_{\rho \bar\mu_2} \Gamma^{\rho}_{\mu_2 \mu_1} + \bar{\chi}^{\bar\mu_1}_{\dot{+}} \tilde\chi^{\mu_2}_{-} k_{\rho \bar\mu_2} \Gamma^{\rho}_{\mu_1 \mu_2} + \bar{\chi}^{\bar\mu_1}_{\dot{-}} \bar{\tilde\chi}^{\bar\mu_2}_{\dot{+}} k_{\mu_1 \bar\rho} \Gamma^{\bar\rho}_{\bar\mu_1 \bar\mu_2} \right]\nonumber \\ 
& & - \bar{F}^{\bar\mu_1} \left[ \tilde\chi^{\mu_2}_{+} \bar{\tilde\chi}^{\bar\mu_2}_{\dot{-}} k_{\mu_2 \bar\rho} \Gamma^{\bar\rho}_{\bar\mu_2 \bar\mu_1} + \chi^{\mu_1}_{+} \bar{\tilde\chi}^{\bar\mu_2}_{\dot{-}} k_{\mu_1 \bar\rho}  \Gamma^{\bar\rho}_{\bar\mu_1 \bar\mu_2} + \chi^{\mu_1}_{-} \tilde\chi^{\mu_2}_{+} k_{\rho \bar\mu_1} \Gamma^{\rho}_{\mu_1 \mu_2} \right]\nonumber \\ 
& & +G^{\mu_2} \left[ \chi^{\mu_1}_{-} \bar\chi^{\bar\mu_1}_{\dot{+}} k_{\rho \bar\mu_1} \Gamma^{\rho}_{\mu_1 \mu_2} +  \bar\chi^{\mu_1}_{\dot{+}} \bar{\tilde\chi}^{\bar\mu_2}_{\dot{-}} k_{\mu_2 \bar\rho} \Gamma^{\bar\rho}_{\bar\mu_2 \bar\mu_1} + \chi^{\mu_1}_{-} \bar{\tilde\chi}^{\bar\mu_2}_{\dot{+}} k_{\rho \bar\mu_2 } \Gamma^{\rho}_{\mu_2 \mu_1}  \right] \nonumber \\ 
& & -  \bar{G}^{\bar\mu_2} \left[ \bar\chi^{\bar\mu_1}_{\dot{-}} \chi^{\mu_1}_{+} k_{\mu_1 \bar\rho} \Gamma^{\bar\rho}_{\bar\mu_1 \bar\mu_2} +  \chi^{\mu_1}_{+} \tilde\chi^{\mu_2}_{-} k_{\rho \bar\mu_2} \Gamma^{\rho}_{\mu_1 \mu_2} + \bar\chi^{\bar\mu_1}_{\dot{-}} \tilde\chi^{\mu_2}_{+} k_{\mu_2 \bar\rho} \Gamma^{\bar\rho}_{\bar\mu_1 \bar\mu_2} \right]\nonumber  
\end{eqnarray}

\begin{eqnarray}
& &+\frac{1}{2} \left[ \tilde\chi^{\mu_2}_{+} \bar{\tilde\chi}^{\bar\mu_2}_{\dot{-}}  \partial_{\mu_2}  \partial_{\bar\mu_2}  +\chi^{\mu_1}_{+} \bar{\tilde\chi}^{\bar\mu_2}_{\dot{-}}  \partial_{\mu_1}  \partial_{\bar\mu_2}  + \chi^{\mu_1}_{-} \tilde\chi^{\mu_2}_{+}  \partial_{\mu_1}  \partial_{\mu_2}  \right]  \bar{\chi}^{\bar\mu_1}_{ \dot{\alpha}} \bar{\chi}^{\bar\nu_1 \dot{\alpha}} \partial_{\bar\mu_1} \partial_{\bar\nu_1} k \nonumber \\
& &-\frac{1}{2} \left[ \bar{\tilde\chi}^{\bar\mu_2}_{\dot{+}} \tilde\chi^{\mu_2}_{-}  \partial_{\mu_2}  \partial_{\bar\mu_2}  +\bar\chi^{\bar\mu_1}_{\dot{+}} \tilde\chi^{\mu_2}_{-}  \partial_{\bar\mu_1}  \partial_{\mu_2}  + \bar\chi^{\bar\mu_1}_{\dot{-}} \bar{\tilde\chi}^{\bar\mu_2}_{\dot{+}}  \partial_{\bar\mu_1}  \partial_{\bar\mu_2}  \right]  \chi^{\mu_1 \alpha} \chi^{\nu_1}_{\alpha} \partial_{\mu_1} \partial_{\nu_1} k \nonumber \\
& &+\frac{1}{2} \left[ \bar\chi^{\bar\mu_1}_{\dot{-}} \chi^{\mu_1}_{+}  \partial_{\mu_1}  \partial_{\bar\mu_1}  +\chi^{\mu_1}_{+} \tilde\chi^{\mu_2}_{-}  \partial_{\mu_1}  \partial_{\mu_2}  + \bar\chi^{\bar\mu_1}_{\dot{-}} \tilde\chi^{\mu_2}_{+}  \partial_{\bar\mu_1}  \partial_{\mu_2}  \right]  \bar{\tilde\chi}^{\bar\mu_2}_{ \dot{\alpha}} \bar{\tilde\chi}^{\bar\nu_2 \dot{\alpha}} \partial_{\bar\mu_2} \partial_{\bar\nu_2} k \nonumber \\
& &-\frac{1}{2} \left[ \chi^{\mu_1}_{-} \bar\chi^{\bar\mu_1}_{\dot{+}}  \partial_{\mu_1}  \partial_{\bar\mu_1}  +\bar\chi^{\bar\mu_1}_{\dot{+}} \bar{\tilde\chi}^{\bar\mu_2}_{\dot{-}}  \partial_{\bar\mu_1}  \partial_{\bar\mu_2}  + \chi^{\mu_1}_{-} \bar{\tilde\chi}^{\bar\mu_2}_{\dot{+}}  \partial_{\mu_1}  \partial_{\bar\mu_2}  \right]  \tilde\chi^{\mu_2 \alpha} \tilde\chi^{\nu_2}_{\alpha} \partial_{\mu_2} \partial_{\nu_2} k \nonumber \\
& &+ \left[ \chi^{\mu_1}_{-} \bar\chi^{\bar\mu_1}_{\dot{-}} {\tilde\chi}^{\mu_2}_{+} \bar{\tilde\chi}^{\bar\mu_2}_{\dot{+}} - \chi^{\mu_1}_{+} \bar\chi^{\bar\mu_1}_{\dot{+}} {\tilde\chi}^{\mu_2}_{-} \bar{\tilde\chi}^{\bar\mu_2}_{\dot{-}} \right] \partial_{\mu_1} \partial_{\bar\mu_1} \partial_{\mu_2} \partial_{\bar\mu_2} k \nonumber \\ 
& & + F^{\mu_1} \bar{F}^{\bar\mu_1} k_{\mu_1 \bar\mu_1} -\frac{1}{2} F^{\mu_1} \bar\chi^{\bar\mu_1}_{\dot{\alpha}} \bar\chi^{\bar\nu_1 \dot\alpha}  k_{\mu_1 \bar\rho} \Gamma^{\bar\rho}_{\bar\mu_1 \bar\nu_1}  - \frac{1}{2} \bar{F}^{\bar\mu_1} \chi^{\mu_1 \alpha} \chi^{\nu_1}_{\alpha} k_{\rho \bar\mu_1} \Gamma^{\rho}_{\mu_1 \nu_1}  \nonumber \\ 
& & + \frac{1}{4} \chi^{\mu_1 \alpha} \chi^{\nu_1}_{\alpha} \bar\chi^{\bar\mu_1}_{\dot{\alpha}} \bar\chi^{\bar\nu_1 \dot\alpha} \partial_{\mu_1} \partial_{\nu_1} \partial_{\bar\mu_1} \partial_{\bar\nu_1} k \nonumber \\ 
& & - G^{\mu_2} \bar{G}^{\bar\mu_2} k_{\mu_2 \bar\mu_2} + \frac{1}{2} G^{\mu_2} \bar{\tilde\chi}^{\bar\mu_2}_{\dot{\alpha}} \bar{\tilde\chi}^{\bar\nu_2 \dot\alpha}  k_{\mu_2 \bar\rho} \Gamma^{\bar\rho}_{\bar\mu_2 \bar\nu_2} + \frac{1}{2} \bar{G}^{\bar\mu_2}  \tilde\chi^{\mu_2 \alpha} \tilde\chi^{\nu_2}_{\alpha} k_{\rho \bar\mu_2} \Gamma^{\rho}_{\mu_2 \nu_2}  \nonumber \\ 
& & - \frac{1}{4} \tilde\chi^{\mu_2 \alpha} \tilde\chi^{\nu_2}_{ \alpha} \bar{\tilde\chi}^{\bar\mu_2}_{\dot{\alpha}} \bar{\tilde\chi}^{\bar\nu_2 \dot\alpha} \partial_{\mu_2} \partial_{\nu_2} \partial_{\bar\mu_2} \partial_{\bar\nu_2} k, \nonumber \\ 
\end{eqnarray}
recall that $k=\sum_{i,j,m,n}\sum_{\underline{\mu},\underline{\nu},\underline{\rho},\underline{\eta}} c_{\underline{\mu},\underline{\nu},\underline{\rho},\underline{\eta}}  P \bar P Q \bar Q$.


\bibliographystyle{utphys}
\bibliography{biblioNatd.bib}

\end{document}